\documentclass[aps,10pt,physrev,superscriptaddress,twocolumn,floatfix]{revtex4-2}

\usepackage[utf8]{inputenc} 
\usepackage[T1]{fontenc}    
\usepackage[colorlinks=true, linkcolor=blue, citecolor=blue, urlcolor=blue]{hyperref}
\usepackage{amsmath}	

\usepackage{tabularx} 

\usepackage{booktabs} 

\usepackage{subcaption}
\usepackage{graphicx} 
\usepackage{xcolor} 

\UseRawInputEncoding

\begin{document}
\sloppy
\renewcommand{\thesection}{\arabic{section}}
\renewcommand{\thesubsection}{\thesection.\arabic{subsection}}
\renewcommand{\thesubsubsection}{\thesubsection.\arabic{subsubsection}}

\title{\textbf{Searching for Binary Black Hole Merger Emission in AGN Disks: Optical and Spectroscopic Follow-up of S240413p }} 


\author{P. Darc}
\email{Contact author: phelipedarc@gmail.com}
\affiliation{Artificial Intelligence for Physics Laboratory (Lab-IA)\\
and Centro Brasileiro de Pesquisas F\'isicas \\
Rua Dr. Xavier Sigaud 150, CEP 22290-180, \\
Rio de Janeiro, RJ, Brazil}
\affiliation{Center for Interdisciplinary Exploration and Research in Astrophysics (CIERA) 
Northwestern University, Evanston, 
IL 60201, USA}
\author{C. R. Bom}

\affiliation{Artificial Intelligence for Physics Laboratory (Lab-IA)\\
and Centro Brasileiro de Pesquisas F\'isicas \\
Rua Dr. Xavier Sigaud 150, CEP 22290-180, \\
Rio de Janeiro, RJ, Brazil}

\author{A. Santos}
\affiliation{Artificial Intelligence for Physics Laboratory (Lab-IA)\\
and Centro Brasileiro de Pesquisas F\'isicas \\
Rua Dr. Xavier Sigaud 150, CEP 22290-180, \\
Rio de Janeiro, RJ, Brazil}

\author{S. Panda}
\thanks{Gemini Science Fellow}
\affiliation{International Gemini Observatory/NSF NOIRLab, Casilla 603, La Serena, Chile}

\author{J. C. Rodr\'iguez-Ram\'irez}
\affiliation{Artificial Intelligence for Physics Laboratory (Lab-IA)\\
and Centro Brasileiro de Pesquisas F\'isicas \\
Rua Dr. Xavier Sigaud 150, CEP 22290-180, \\
Rio de Janeiro, RJ, Brazil}

\author{C. D. Kilpatrick}
\affiliation{Center for Interdisciplinary Exploration and Research in Astrophysics (CIERA) 
Northwestern University, Evanston, 
IL 60201, USA}

\author{C. Mendes de Oliveira} 
\affiliation{Universidade de S\~ao Paulo, IAG, Rua do Mat\~ao 1225, S\~ao
Paulo, SP, Brazil}

\author{A. Kanaan}
\affiliation{Departamento de F\'isica, \\ Universidade Federal de Santa Catarina, Florian\'opolis, SC, 88040-900, Brazil}

\author{T. Ribeiro}
\affiliation{Departamento de Astronomia, Instituto de F\'isica, Universidade Federal do Rio Grande do Sul (UFRGS), Av. Bento Goncalves 9500.}

\author{W. Schoenell}
\affiliation{NOAO, P.O. Box 26732, Tucson, AZ 85726}

\begin{abstract}
The conditions under which binary black hole (BBH) mergers embedded in active galactic nucleus (AGN) disks produce detectable optical counterparts remain poorly constrained observationally. We report multi-epoch optical imaging and spectroscopic follow-up of S240413p, an O4 BBH candidate with 98\% classification confidence and one of the smallest sky localizations reported to date ($\sim$$61~\mathrm{deg}^2$), obtained with the T80-South telescope through the S-PLUS Transient Extension Program (STEP). Our observations cover the 99\% credible region across epochs that span $\sim$300 days post-merger. We prioritize AGN-hosted environments and identify two transient candidates, STEP2024gab/ZTF18acvgziq and STEP2024phe/ZTF19aaflhnr. SOAR/Goodman spectroscopy and archival DESI spectra yield host supermassive black hole masses of $\log M_\mathrm{SMBH}/\mathrm{M}_\odot = 7.15 \pm 0.05$ and $8.02 \pm 0.04$. We compute predicted flare delay distributions for each host using a thermal radiation-driven outflow emission model and the spectroscopically derived host properties. Migration traps produced by thermal torques occur at $R_\text{{BH}}/R_g \approx 10^{4.2}$ and $10^{3.4}$ for the two hosts, with predicted flare delays spanning tens to several hundred days; our late epoch at $\sim$ 300 days coincides with both the peak of these distributions and the migration trap locations, while early epochs overlap only their tails. An independent five-dimensional detection efficiency analysis using BBH light-curves and \texttt{Teglon} confirms that $M_\mathrm{SMBH} \sim 10^7$--$10^8~M_\odot$ AGN environments at merger distances of $\sim$0.003--0.012~pc are the most favorable for detectable emission, a regime both candidate hosts occupy. We find no confirmed counterpart; a seasonal visibility gap leaves open the possibility that a flare occurred undetected, the merger may not have occurred within the AGN disk itself, or any emission may have been obscured by intrinsic AGN variability. These results demonstrate that long-baseline, AGN-prioritized monitoring is a necessary condition for accessing the highest-probability region of BBH merger parameter space, and establish the need for physically informed follow-up strategies in the Rubin/LSST era.
\end{abstract}


\maketitle

\section{Introduction}\label{sec:introduction}

The direct detection of gravitational waves (GWs) by the LIGO/Virgo/KAGRA (LVK) Collaborations \citep{Aasi_2015,Acernese_2015} has revolutionized astrophysics, providing a unique probe of compact-object mergers and fundamental cosmic parameters. The first detection, GW150914, observed in 2015 by the Advanced LIGO detectors, confirmed the existence of binary black hole (BBH) mergers and marked the beginning of gravitational-wave astronomy \citep{Abbott2016}. Since then, the expansion of the global interferometer network, including Advanced Virgo and KAGRA, has increased both the detection rate and sensitivity, enabling the characterization of hundreds of BBH systems, two binary neutron star (BNS) mergers, and a few neutron star-black hole (NSBH) events across multiple observing runs \citep{GWTC1,GWTC2,GWTC3,GWTC4a}.

The fourth observing run (O4), initiated in 2023, has already delivered hundreds of candidate detections with improved localization regions and higher signal-to-noise ratios, reflecting the rapid maturation of GW detection capabilities and the transition of GW astronomy into a precision, population-scale science \citep{GWTC4a}. One of the most promising techniques to emerge from this field is the standard siren method, first proposed by \cite{Schutz1986}, which utilizes the luminosity distance directly inferred from GW strain amplitude measurements. When combined with an independent redshift estimate, typically obtained through the identification of a host galaxy or an electromagnetic (EM) counterpart, this approach allows cosmological parameters such as the Hubble constant ($H_0$) to be inferred via the distance-redshift relation.

The vast majority of GW detections to date arise from BBH mergers.
In standard formation channels, BBH mergers are expected to be electromagnetically dark, as the lack of surrounding material prevents accretion or shocks capable of powering a detectable electromagnetic counterpart. However, recent studies have suggested that a significant fraction -- possibly $20$--$80\%$ \citep{ratebbhFord2022} -- of the BBH merger rate measured by LIGO/Virgo, estimated at $\sim 24\,\mathrm{Gpc}^{-3}\,\mathrm{yr}^{-1}$, may occur within the accretion disks of active galactic nuclei (AGNs) \citep{ratebbhArcaSedda2023,ratebbhGrbner2020,fractionagn_Zhu2025}.

The gas-rich environment of an AGN disk provides an opportunity for the embedded BBHs to generate electromagnetic emission through shock heating, disk perturbations, or jet formation. In particular, asymmetric BBH mergers (in mass and/or spin) will receive a recoil velocity (or kick) upon merger due to anisotropic gravitational-wave emission \citep{Lousto_2010}. The interaction between the BH and the surrounding AGN disk can generate luminous shocks and accretion-driven flares \citep{McKernan2019,Kimura2021,juan_publicado, tagawa_phe_model}. These flares, often referred to as \emph{dark flares} \citep{Darc25_multimomel_constraint_bbh}, differ from variability driven by the typical stochastic processes in active galactic nuclei. Depending on the disk density and kick velocity, such events can produce optical and/or UV afterglows lasting from days to weeks, thereby providing a potential electromagnetic signature of BBH mergers.

Several AGN flares have been proposed as tentative GW counterparts, including ZTF19abanrhr, which was spatially coincident with the GW190521 event \citep{Graham2020}. However, intrinsic stochastic variability of AGNs complicates the establishment of a robust causal connection, especially given the large GW localization regions and the possibility of multiple flaring events occurring within the same spatial and temporal window \citep{Palmese_2021bbh}. For example, by cross-matching all GWTC-4.0 events with AGN flares identified over six years of Zwicky Transient Facility (ZTF) DR23 observations, \citet{fractionagn_Zhu_he2026} inferred that $7^{+24}_{-5}\%$ of BBH mergers may be associated with a ZTF flare.

Collectively, these findings underscore the growing importance of multiwavelength and multimessenger follow-up campaigns. Coordinated wide-field optical surveys with rapid spectroscopic capabilities -- such as S-PLUS/STEP \citep{Santos24} and SOAR/Goodman \footnote{\url{https://noirlab.edu/science/programs/ctio/telescopes/soar-telescope}} \citep{Clemens2004} -- are particularly well suited to identify, characterize, and constrain the electromagnetic counterparts of both BNS and BBH mergers, thereby advancing our understanding of compact-object astrophysics and refining measurements of fundamental cosmological parameters.

During the fourth observing run (O4), the LIGO/Virgo Collaboration reported the gravitational-wave superevent S240413p, detected on 2024-04-13 at 02:20:19 UTC (MJD 60413.09884). This event is notable for its well-constrained 50\% sky localization area of approximately $11\,\mathrm{deg}^2$ and its low false-alarm rate of $\sim 3.2 \times 10^{-10}\,\mathrm{Hz}$.  By cross-matching the three-dimensional localization volume with the Million Quasars catalog \citep[Milliquas;][]{milliquas2023yCat.7294....0F}, we identified 168 AGN candidates within the 90\% credible region. Two of these (STEP2024gab and STEP2024phe candidates) exhibited positive residual flux in our initial search, making them plausible hosts for a binary black hole merger with a potentially detectable electromagnetic counterpart.

In this paper, we present the optical follow-up of the GW superevent S240413p conducted with the T80-South telescope. We describe the observing strategy, the candidate-vetting procedure, and the spectroscopic characterization of two AGN candidates located within the three-dimensional localization volume. Using the inferred properties of their supermassive black holes (SMBH), we investigate the physical conditions under which an electromagnetic counterpart to a BBH merger could arise in such environments. We further employ the observational limiting magnitudes to constrain the region of parameter space in the emission scenario proposed by \citeauthor{juan_novo} that would yield a detectable counterpart during our search.

This paper is organized as follows. In Section~2, we describe the optical imaging and spectroscopic data acquisition and reduction. Section~3 presents the spectral decomposition, light-curve analysis, and the methodology used to constrain the dark-flare parameters and their observational properties, together with the BBH emission model adopted in this work. Section~4 presents the resulting constraints and discusses their implications, and Section~5 summarizes our conclusions. Throughout this paper, we assume a standard $\Lambda$CDM cosmology with $H_0 = 70\,\mathrm{km\,s^{-1}\,Mpc^{-1}}$ and $\Omega_\mathrm{m} = 0.315$.

\begin{table*}[htp]
\centering
\caption{Summary of T80S Follow-up Observations of S240413p}
\label{tab:t80s_followup}
\begin{tabular}{l|c|c|c|c}   
\hline
Date & Filter & Exposure Time (s) & Days after Merger & $3\sigma$ Limiting Mag (AB) \\
\hline
2024-04-16 & $g$ & 120 & 3   & 21.6 \\
2024-04-16 & $r$ & 150 & 3   & 21.7 \\
2024-04-25 & $i$ & 240 & 12  & 21.5 \\
2024-05-11 & $i$ & 200 & 28  & 21.3 \\
2025-04-13 & $i$ & 200 & 366 & 20.8 \\
2025-04-18 & $i$ & 200 & 371 & 21.2 \\
2025-05-23 & $i$ & 200 & 406 & 21.7 \\
2025-05-24 & $i$ & 200 & 407 & 21.5 \\
\hline
\end{tabular}

\end{table*}

\section{Observations and Data Analysis}\label{sec:data}

\subsection{GW Data: The LIGO/Virgo Event S240413p}
The gravitational wave superevent S240413p was reported by the LVK alert stream during real-time data processing from the LIGO Hanford (H1), LIGO Livingston (L1), and Virgo Observatory (V1) detectors at 2024-04-13 02:20:19 UTC (MJD: 60413.09884259259). The gravitational wave superevent designated S240413p was identified by the CWB \citep{klimenko2016}, {\tt GstLAL} \citep{messick2017,tsukada2023}, {\tt MBTA} \citep{aubin2021}, and {\tt PyCBC} \citep{dalCanton2021} live analysis pipelines. After parameter estimation by RapidPE-RFIT \citep{pankow2015, jacob2018}, the updated classification on the origin of the GW signal was 98\% BBH, 2\% Terrestrial, 0\% NSBH, and 0\% for BNS, with a False Alarm Rate (FAR) of $\sim 3.2 \times 10^{-10}\,\mathrm{Hz}$ ($\sim$1 per 100 years), estimated by {\tt pyCBC}. The luminosity distance was $526 \pm 101\,\mathrm{Mpc}$, corresponding to a redshift of 0.11 $\pm$ 0.02, and a sky area in the 50\% (99\%) credible region of $11\,\mathrm{deg}^2$ ($61\,\mathrm{deg}^2$). 

\subsection{T80-South Observations and Data Analysis}
We triggered Target-of-Opportunity (ToO) observations under the S-PLUS Transient Extension Program (STEP; \citealt{Santos24}), interrupting the Main Survey of the Southern Photometric Local Universe Survey \citep[S-PLUS;][]{MendesdeOliveira2019}. S240413p satisfied our selection criteria for well-localized events, with a 90\% credible region of less than $100~deg^2$ and a FAR of $\sim$$10^{-10}$~Hz, corresponding to less than one event per century. Although a small terrestrial probability is reported, the precise sky localization and low FAR across all three detectors motivate a dedicated electromagnetic follow-up campaign. 

Our follow-up strategy is motivated primarily by optical and UV emission models \citep{McKernan2019, juan_publicado, tagawa_phe_model, RodrguezRamrez2025}, which predict two broad classes of counterparts. Early-type flares peak within $\lesssim$50~days of the merger and last days to weeks, while late-type flares peak between $\sim$50 and $\sim$450~days post-merger with durations that can exceed $\sim$200~days. This bimodal delay structure reflects the range of possible merger radii within the AGN disk and the diversity of outflow propagation timescales. Guided by these predictions, we divided our follow-up campaign into early- and late-time observing phases.

\subsubsection{Observation Strategy and Initial Search}

Unfavorable weather conditions during the first two nights following the gravitational-wave merger delayed the start of our observations. The earliest epoch was obtained 3 days after the merger, on 2024-04-16 at 01:27~UTC using the T80-Cam instrument mounted on the T80-South (T80S) telescope at the Cerro Tololo Inter-American Observatory (CTIO). T80S is a 0.8-m robotic telescope with a field of view of approximately $2\,\mathrm{deg}^2$. The instrument is equipped with a 12-band optical filter system consisting of five broad bands ($u,g,r,i,z$) and seven narrow bands (J0378, J0395, J0410, J0430, J0515, J0660, J0861), designed to sample key spectral features such as [O\,II], H$\alpha$, and the Ca triplet.

We partitioned the updated sky map into $2\,\mathrm{deg}^2$ tiles, matching the T80S field of view. Owing to the relatively small localization area, we adopted a direct tiling strategy based on sky-map probability and observability constraints, selecting fields with an altitude greater than $40^\circ$ during the available observing windows. Rather than reweighting the localization probability using galaxy or AGN catalogs, as done in previous searches \citep{Darc25_multimomel_constraint_bbh}, we prioritized tiles using the raw sky-map probability combined with visibility criteria. We observed the sky map using 120\,s exposures in the $g$ band for the first pass over all selected tiles, followed by a second independent pass in the $r$ band with 150\,s exposures, totaling 56 pointings on the first night. These observations reached a mean $3\sigma$ limiting magnitudes of 21.6~mag (AB) in $g$ and 21.7~mag (AB) in $r$. Overall, we covered 47 of the $61\,\mathrm{deg}^2$ contained within the 99\% credible region of the localization map. The choice of filters and exposure times was optimized to maximize the probability of detecting a counterpart to the GW event while maintaining an efficient balance between survey depth, sky coverage, and telescope time usage under Target-of-Opportunity constraints.


Following the initial search within the S240413p localization region, we processed the data with our difference-imaging pipeline and candidate-selection procedure (described in the next section). This search yielded 242 transient candidates. We then cross-matched these candidates with the Million Quasars catalog \citep[Milliquas;][]{milliquas2023yCat.7294....0F}, the largest currently available compilation of AGNs, containing over one million AGN-like sources. This cross-match identified two likely host candidates associated with significant difference-imaging detections in our pipeline, hereafter designated STEP2024gab and STEP2024phe. 

Motivated by the spatial coincidence of these two candidates with known AGNs, we triggered spectroscopic follow-up observations with the SOAR/Goodman instrument for both targets. In addition to the initial search, we conducted further imaging campaigns over a subset of tiles covering $26\,\mathrm{deg}^2$ within the credible region at 12 and 28 days after the merger. For these epochs, we used the $i$-band filter with single-pass exposures of 240\,s and 200\,s per tile, optimized to increase sensitivity to faint emission emerging on timescales of days to weeks after the merger. The resulting subtracted images reached mean $3\sigma$ limiting magnitudes of 21.5 and 21.3~mag (AB), respectively.

\begin{figure}[h!]
    \centering
    \includegraphics[width=0.8\linewidth]{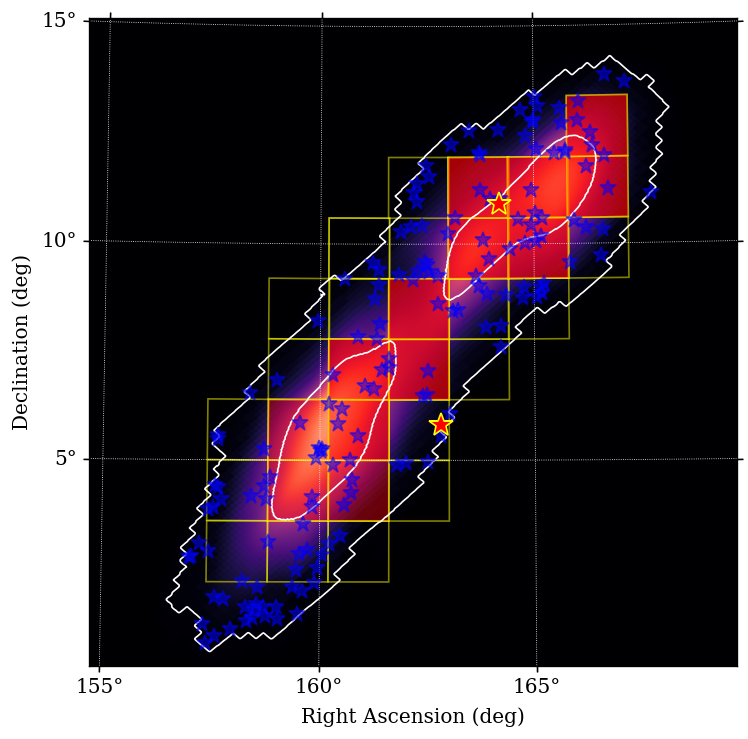}
     \caption{Sky localization of the GW superevent S240413p reported by the LVK Collaboration. The probability map is shown in celestial coordinates, with contours enclosing the 50\% ($11\,\mathrm{deg}^{2}$) and 90\% ($34\,\mathrm{deg}^{2}$) credible regions. The initial T80S tiling is indicated by orange squares, while fields observed in subsequent follow-up are shown as filled orange squares. The campaign covered $47\,\mathrm{deg}^{2}$ of the 99\% credible region of the GW localization. AGN candidates cross-matched with the Milliquas catalog are marked as blue stars. The two red stars indicate AGN candidates exhibiting positive residual flux in the initial difference-imaging search, labeled STEP2024gab (upper source) and STEP2024phe (lower source).}
    \label{fig:skymap}
\end{figure}

Our late-time follow-up campaign began approximately one year after the GW event, with observations obtained on 2025-04-13, 2025-04-18, 2025-05-23, and 2025-05-24. These epochs used the $i$-band filter with 200\,s exposures in a single pass over the selected subset of tiles. The subtracted images reached median $3\sigma$ limiting magnitudes of 20.8, 21.2, 21.7, and 21.5~mag (AB) for each respective night. All epochs were processed through the same vetting pipeline used in the initial search, and no additional viable candidates were identified. A summary of the photometric observation properties conducted by the STEP program is presented in Table~\ref{tab:t80s_followup}. Figure~\ref{fig:skymap} shows our tiling coverage over the S240413p localization probability region for the initial search (orange tiles) and subsequent
follow-up (filled orange tiles). The two targets associated with AGNs are marked with red stars.

\subsubsection{S-PLUS Transient Extension Pipeline (STEP)}
The S-PLUS Transient Extension Pipeline (STEP) is the time-domain branch of the ongoing S-PLUS survey, designed for systematic transient discovery and rapid multi-messenger follow-up. STEP operates at the Laborat\'orio de Intelig\^encia Artificial \footnote{\href{https://labia.cbpf.br/}{https://labia.cbpf.br/}} (LabIA) of the Centro Brasileiro de Pesquisas F\'isicas (CBPF), where data reduction, difference imaging, and candidate vetting are performed in near real time. The pipeline is optimized for wide-field optical searches with the T80S telescope and supports both untargeted survey operations and rapid-response observations to gravitational-wave and other high-energy transient alerts. STEP currently operates in two complementary configurations:

(i) Nightly scheduled searches. In this mode, STEP processes images acquired during regular S-PLUS observations from the most recent nights. The data are automatically reduced, astrometrically refined, and subtracted against external templates to identify new transient candidates within the S-PLUS footprint. This configuration includes both blind searches in newly observed survey fields and monitoring of previously reported transients, particularly supernova candidates registered in the Transient Name Server (TNS), internally designated as ``SN fields.'' This operational mode provides continuous coverage of the southern sky and generates a steady stream of transient discoveries and light-curve follow-up.

(ii) Target-of-Opportunity (ToO) mode. In response to external alerts, STEP switches to a rapid ToO configuration. In this mode, images are acquired on-the-fly and undergo expedited instrumental calibration, including bias and flat-field corrections, astrometric solution, and immediate difference imaging. The first reduced products are typically available within minutes of data acquisition, enabling prompt transient identification and reporting. This configuration is designed to support follow-up observations of gravitational-wave candidates, gamma-ray bursts (GRBs), and, more recently, fast X-ray transients (FXTs) distributed through the General Coordinates Network (GCN).

This dual-mode operational strategy allows STEP to function both as a wide-field survey transient program and as a rapid-response facility for multi-messenger astronomy and other reported transients in GCNs. The T80S images are processed with the STEP pipeline, a dedicated data reduction and difference imaging framework developed for transient detection. STEP is built on top of {\tt photpipe} \citep{Rest2005} and tailored to the T80Cam instrument characteristics, including its pixel scale, wide field of view, and survey observing strategy, using Dark Energy Survey \citep{des_white_paper} or PAN-STARRS DR2 as template images \citep{PanSTARRSDR2}.

\subsubsection{Data Reduction and Image Subtraction}

As detailed in \citep{Santos24,Darc25_multimomel_constraint_bbh}, the imaging processing pipeline processes bias- and flat-field-corrected images with an initial astrometric solution. The astrometry is refined using the {\tt msccmatch} task in {\tt IRAF} \citep{IRAF}, and both science and template images are resampled onto a common reference grid with {\tt SWarp} \citep{Bertin2010}. A photometric zero-point is then derived for subsequent magnitude measurements.

Image subtraction is performed using the algorithm of \citet{AlardLupton1998}, implemented through {\tt hotpants} \citep{Becker2015}. Forced photometry is carried out with a customized version of {\tt DoPhot} \citep{Schechter1993}. Transient candidates are identified as sources detected in the subtracted images with ${\rm S/N} > 5$ above the local background based on PSF photometry, and are subsequently subjected to additional vetting procedures.

\subsubsection{Candidate Vetting Stage}

The candidate vetting procedure for this event follows methods similar to those described by \citet{Kilpatrick_2021_gravcole} and \citet{Darc25_multimomel_constraint_bbh}. All detected candidates are first cross-matched against the Gaia DR3 catalog of variable stars, and sources with counterparts within a 1~arcsecond radius are removed.

To further reduce the number of candidates requiring visual inspection, we apply a convolutional neural network (CNN) classifier based on the MobileNet architecture, trained to distinguish genuine astrophysical transients from image artifacts \citep[see][]{Santos24}. After this automated classification stage, we impose a signal-to-noise threshold of SNR~$\geq 10$ to retain only statistically significant detections.

The remaining candidates are visually inspected and cross-matched with the Minor Planet Center database within a 10~arcsecond radius to reject known solar system objects. After the first night of observations, two promising candidates remained within the $99\%$ credible region: STEP2024gab and STEP2024phe.

\subsection{Spectroscopic Observations}

\subsubsection{SOAR/Goodman}

Spectroscopic follow-up observations were obtained with the Goodman High Throughput Spectrograph (HTS) mounted on the 4.1-m Southern Astrophysical Research Telescope (SOAR) \footnote{Based in part on observations obtained at the Southern Astrophysical Research (SOAR) telescope, a joint project of the Minist\'erio da Ci\^encia, Tecnologia e Inova\c{c}\~oes (MCTI/LNA) of Brazil, the U.S. National Science Foundation’s NOIRLab, the University of North Carolina at Chapel Hill (UNC), and Michigan State University (MSU).}. This program was awarded ToO observing time to follow up optical candidates associated with gravitational-wave events discovered during the O4 observing run of the LIGO/Virgo/KAGRA collaboration. The rapid-response ToO mode is particularly important for identifying transient spectroscopic signatures from short-lived counterparts such as the hypothesized ``dark flares''.

The optical spectrum of ZTF19aaflhnr (STEP2024phe) was obtained on 2024 March 15 under program SO2024A-022 (PI: De Bom \& Kilpatrick). Observations were performed with the Goodman spectrograph in the Red Camera configuration using the 400 SYZY grating, providing a resolving power of $R \sim 2200$ at 5000~\AA\ and covering a wavelength range of $4951$--$8980$~\AA. The data were acquired using a $1.0^{\prime\prime}$ long slit in two exposures of 600~s each. The detector was operated with a read noise of 3.89~$e^{-}$ and a gain of 1.48~$e^{-}$/ADU in $2\times2$ binning mode. Observations were conducted at an average airmass of 1.25 under stable atmospheric conditions.

The data reduction was performed using the {\tt PypeIt} pipeline \citep{Prochaska2020a, Prochaska2020b}. Standard long-slit reduction procedures were applied, including bias subtraction, flat-field correction, cosmic-ray rejection, wavelength calibration, sky subtraction, and one-dimensional spectral extraction. Wavelength calibration was obtained using HgArNe arc-lamp exposures taken immediately after the science frames, achieving an RMS accuracy of $\sim0.25$~\AA. Flux calibration was performed using spectrophotometric standard stars, and telluric absorption corrections were applied using a standard telluric grid. The object spectrum was extracted using the optimal object finding algorithm from {\tt pypeit} based on the standard Horne algorithm \citep{Prochaska2020b}.

Due to unfavorable atmospheric conditions during the SOAR observations, the spectrum obtained for the ZTF18acvgziq (STEP2024gab) candidate has a very low signal-to-noise ratio and is not suitable for scientific analysis. We therefore supplemented our dataset with archival spectroscopic observations from the Sloan Digital Sky Survey (SDSS; \citep{SDSS_Almeida2023}) and the Dark Energy Spectroscopic Instrument (DESI; \citep{DESICollaboration2022}) archival spectroscopy.

\subsubsection{SDSS and DESI}

We complemented our spectroscopic analysis with archival data from SDSS and DESI for both targets. Archival SDSS spectra were retrieved from the SDSS Science Archive Server (SAS) database \citep{SDSS_Almeida2023,SDSSBlanton2017}, which provides access to calibrated spectra and imaging products, along with tools for interactive visualization and download. DESI spectra were obtained by cross-matching our targets with the publicly available DESI DR1 catalog \citep{DESIDR1} and querying the corresponding spectral products through the SPARCL database \citep{SPARCL_2024arXiv240105576J}.

ZTF18acvgziq (STEP2024gab) has both SDSS and DESI archival spectroscopy available. The SDSS spectrum corresponds to source SDSS J105644.96+105455.7 and was obtained on 22 April 2004 (MJD 53117). A DESI spectrum is also available for DESI J164.1873+10.9154, observed on 01 December 2021 (MJD 59549). For the second target, ZTF19aaflhnr (STEP2024phe), an archival SDSS spectrum is available from observations obtained on 31 January 2003 (MJD 52670). No DESI EDR/DR1 spectrum is currently available for this source.

\section{Analysis and Discussion}\label{sec:methods}

\subsection{Spectral decomposition using {\tt PyQSOFit}}

We perform spectral decomposition of each AGN spectrum using {\tt PyQSOFit} \citep{Guo_2018ascl.soft09008G}. The spectra are first corrected to the rest frame and for Galactic extinction, adopting the extinction curve of \cite{Cardelli_1989ApJ...345..245C} and the dust reddening map of \cite{Schlegel_1998ApJ...500..525S}. Host-galaxy decomposition is then carried out with galaxy eigenspectra from \cite{Yipa_2004AJ....128..585Y} and quasar eigenspectra from \cite{Yipb_2004AJ....128.2603Y}, as implemented in {\tt PyQSOFit}. For the continuum modeling, we fit a power law, optical Fe {\sc ii}, and Balmer continuum components, using continuum windows defined in \cite{Rakshit_2020ApJS..249...17R} and \cite{Panda_2024ApJS..272...13P}. The optical Fe {\sc ii} emission is modeled with the empirical template of \cite{Boroson_1992ApJS...80..109B}, covering 3686-7484 \AA, with normalization, broadening, and wavelength shift as free parameters. Emission lines are then fitted with Gaussian profiles following \cite{Shen_2019ApJS..241...34S} and \cite{Rakshit_2020ApJS..249...17R}. Depending on the redshift and spectral coverage, the fitted lines include: broad and narrow H$\alpha$$\lambda$6562, [N {\sc ii}]$\lambda$$\lambda$6549,6585, [S {\sc ii}]$\lambda$$\lambda$6718,6732, broad and narrow H$\beta$$\lambda$4861, and [O {\sc iii}]$\lambda$$\lambda$5007,4959. All fits are run with Monte Carlo simulations based on the observed spectral error array, which propagates uncertainties into the decomposition results. The host-galaxy fits in {\tt PyQSOFit} are restricted to rest-frame 3450-8000 \AA. Each line complex (H$\beta$ and H$\alpha$ in this work) is fitted within relatively narrow spectral windows ($\sim$100-150 \AA; see the lower panels of Figure 1), after subtracting the power-law continuum and host contribution, leaving only the emission-line profiles to be modeled.

The results of the spectral decomposition are presented in Figures \ref{fig:spec_sdss_ztf18} and \ref{fig:spec_desi_ztf18} (for ZTF18acvgziq/STEP2024gab) and Figures \ref{fig:spec_sdss_ztf19} and \ref{fig:spec_soar_ztf19} (for ZTF19aaflhnr/STEP2024phe), with the derived spectral properties listed in Tables \ref{tab:spec-table1} and \ref{tab:spec-table2}.

\subsubsection{STEP2024gab/ZTF18acvgziq}

\begin{figure*}[tp]
    \centering
    \includegraphics[width=\linewidth]{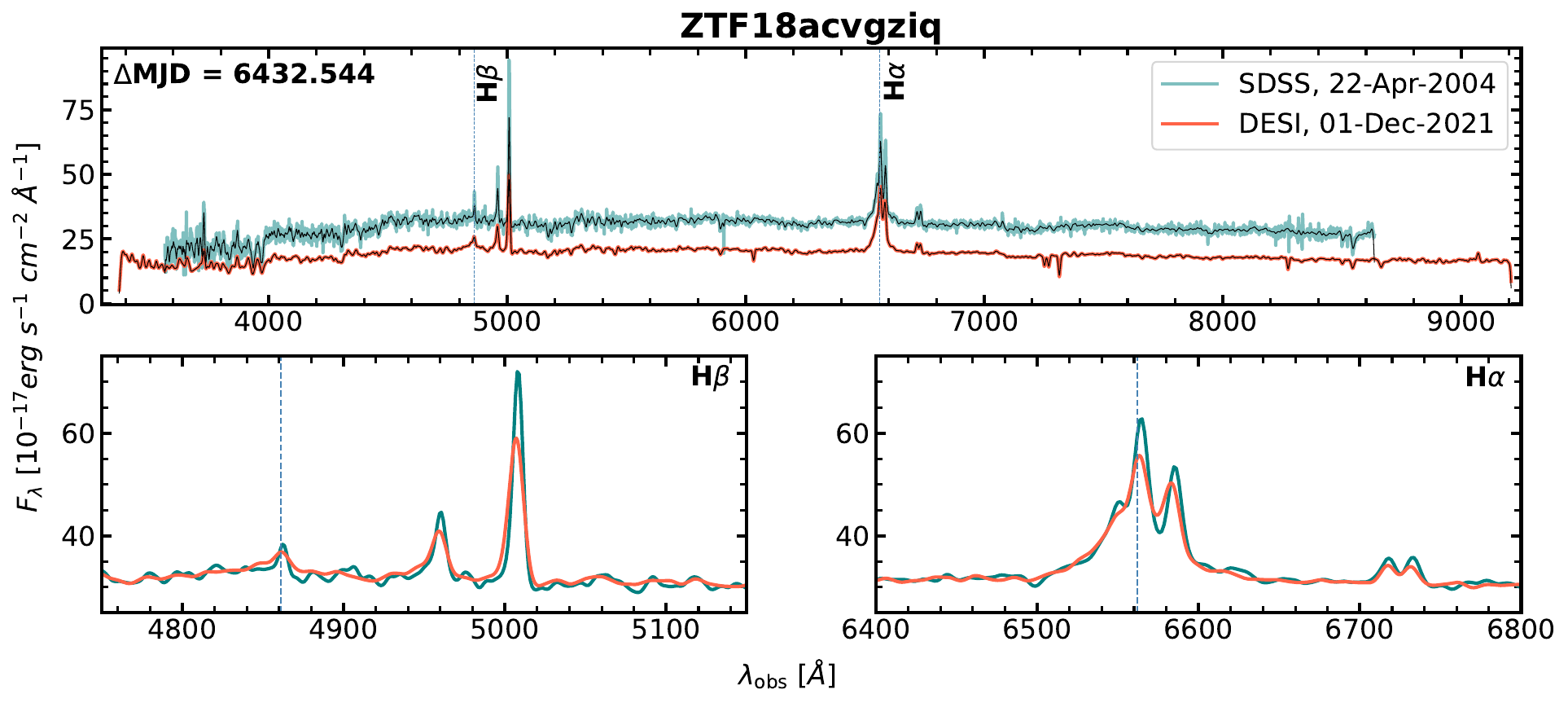}
    \caption{Alt name: \textbf{2MASS J10564497+1054558}, in lower panels, the DESI spectrum is shifted to match the continuum level for the SDSS (older) spectrum - to allow comparison of the change in the respective line profiles.}
    \label{fig:spec_ztf18}
\end{figure*}

\begin{figure*}[tp]
    \centering
    \includegraphics[width=\linewidth]{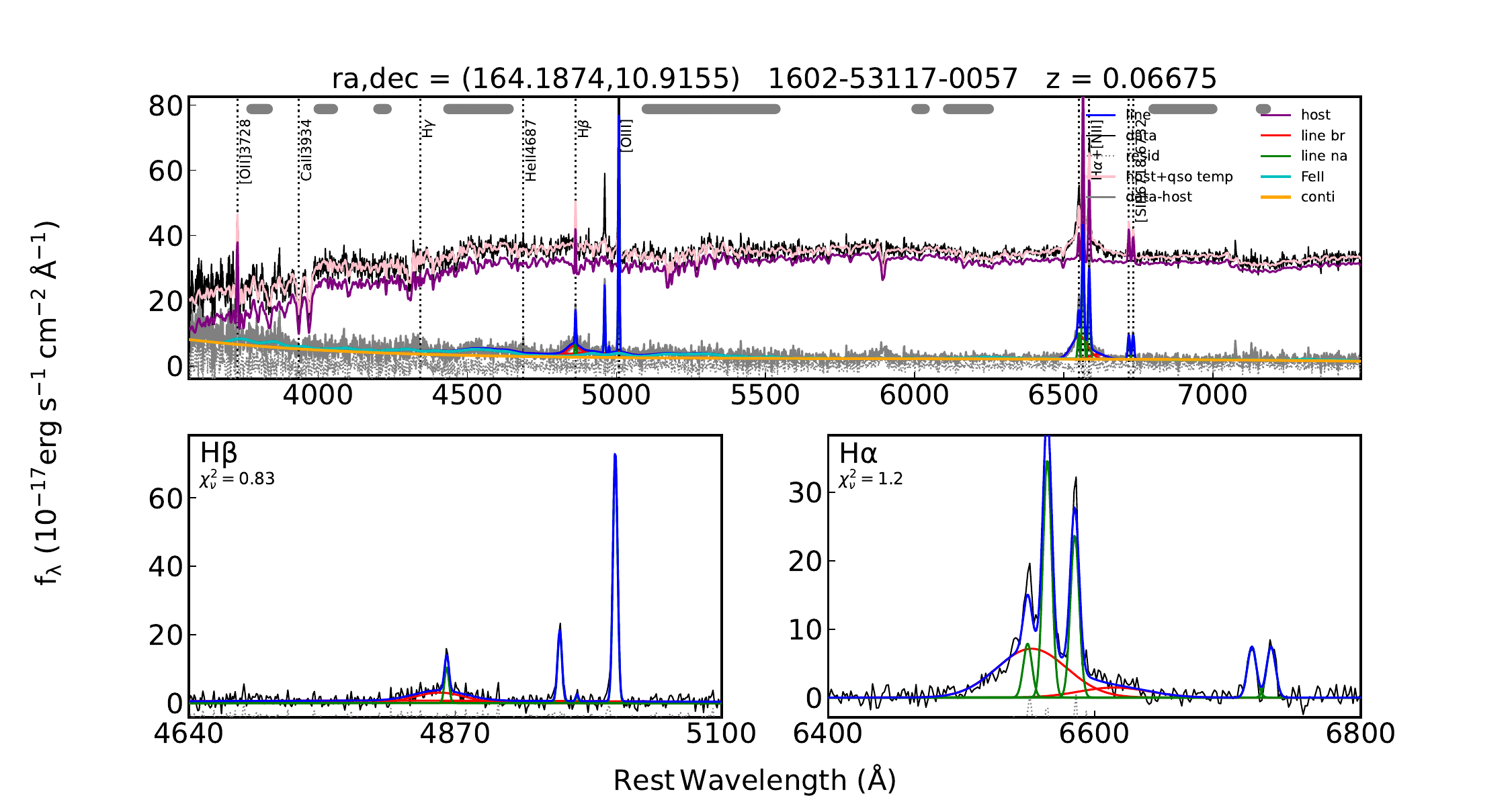}
    \caption{Exemplary fit using {\tt PyQSOFit} \citep{Guo_2018ascl.soft09008G} for a quasar spectrum (ZTF18acvgziq, SDSS). We show the SDSS spectrum (black), power-law continuum (yellow), Fe {\sc ii} pseudocontinuum (in addition to the power-law continuum; light green), broad emission lines (red), narrow emission lines (dark green), and the total best-fit QSO model (blue), which is the sum of continuum and emission lines. The host galaxy contribution is shown in magenta, while the host-subtracted data are shown with a continuous gray line, and the sum of the host and QSO model is shown in pink. Top panel: the rest-frame central wavelengths for prominent emission lines are shown using the dashed vertical lines. The sky coordinates (in degrees) and the redshift for the sources are quoted in the title of the figure. Bottom panels: a zoomed-in version of individual line complexes (left: H$\beta$, and right: H$\alpha$). The residuals are shown in dotted gray in each panel.}
    \label{fig:spec_sdss_ztf18}
\end{figure*}

\begin{figure*}[tp]
    \centering
    \includegraphics[width=\linewidth]{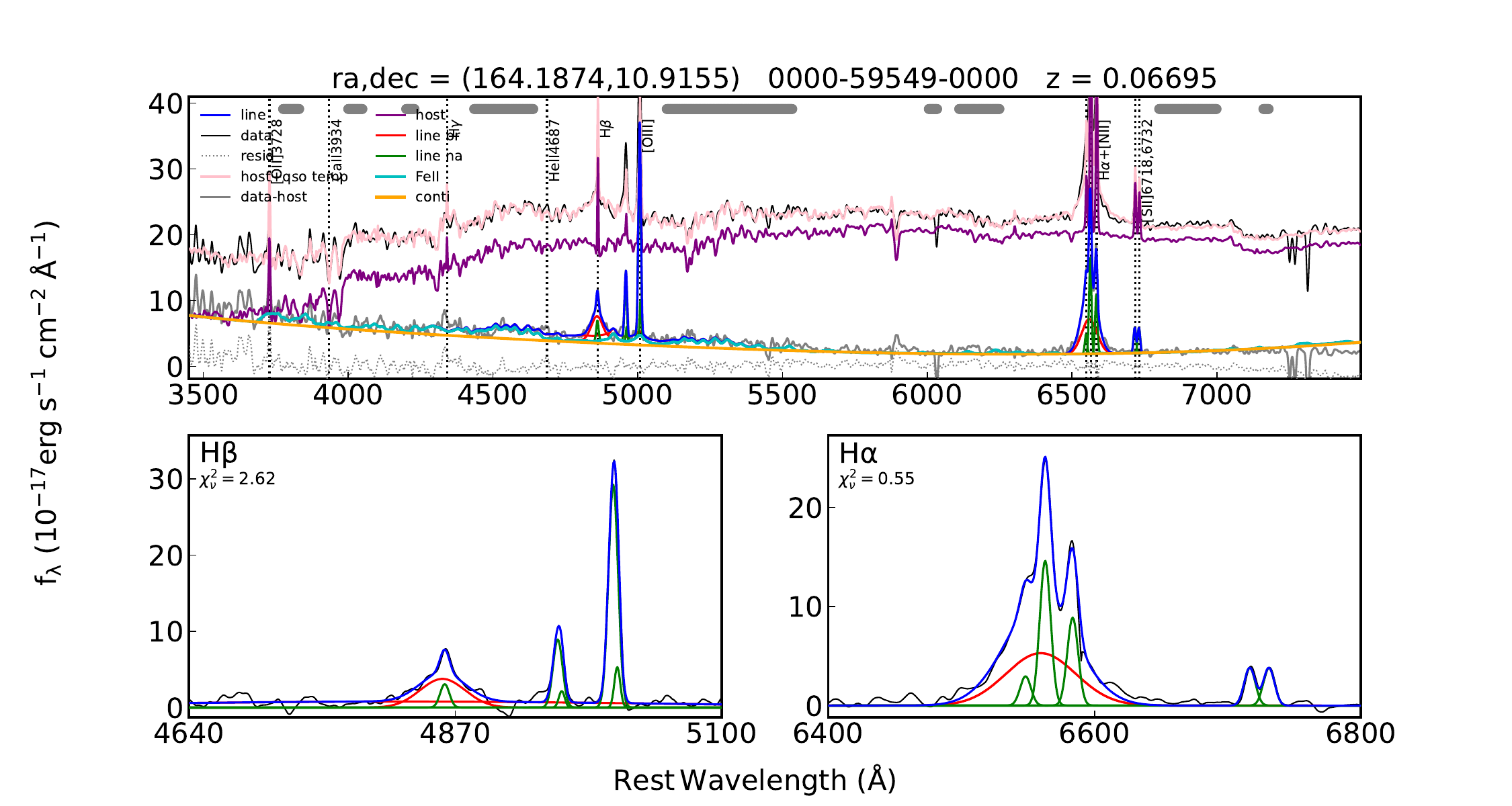}
    \caption{Similar to Figure \ref{fig:spec_sdss_ztf18} but for the more recent DESI spectrum for the same source.}
    \label{fig:spec_desi_ztf18}
\end{figure*}

The first target, ZTF18acvgziq, has both SDSS and DESI archival spectroscopy available: an SDSS spectrum of SDSS J105644.96+105455.7, and a DESI spectrum of DESI J164.1873+10.9154. These two spectroscopic epochs are separated by nearly 18 years. The corresponding spectra are shown in Figure~\ref{fig:spec_ztf18}. The SDSS spectrum, i.e., the earlier epoch (marked in teal), demonstrates a higher total flux level - dominated by the host galaxy flux that contributes to 88.6\% of the total flux computed at 5100 \AA. This results in an AGN continuum luminosity (after reddening correction and host subtraction) log L$_{\rm 5100}$ = 42.132$\pm$0.012 erg s$^{-1}$. On the other hand, the DESI spectrum (marked in orange) shows a much shallower gradient of flux change, especially in the shorter wavelength region - a sign that the spectrum has a reduced contribution from the host galaxy (69.4\%). The corresponding log L$_{\rm 5100}$ = 42.521$\pm$0.017 erg s$^{-1}$, that is, $\sim$2.5 times higher than the SDSS epoch. In the bottom panel of the same figure, we highlight the changes in the emission line profiles for H$\beta$ (left) and H$\alpha$ (right) complexes for the two spectral epochs. For clarity, we scaled the DESI spectrum to match the continuum level of the SDSS spectrum while showing these line complexes, in addition to rebinning the spectra with 2\AA\ bins. Contrary to an AGN-dominated case, here we see that the emission lines have weaker line flux contributions for the more recent spectral epoch (DESI), consistently for both the Balmer lines relative to the older epoch (SDSS). In addition, we notice that the [O {\sc iii}] doublet shows a slight blueward asymmetry for the DESI spectrum, perhaps reminiscent of an outflowing wind associated with the outer edge of the broad-line region (BLR). These inferences are reaffirmed after the spectral decomposition (see Figures \ref{fig:spec_sdss_ztf18} and \ref{fig:spec_desi_ztf18}).

To estimate the black hole mass and the accretion rate for the source, we utilize the prescription from \cite{Panda_MQS_2024ApJS..272...11P} based on the earlier works of \cite{Shen_2011ApJS..194...45S} and \cite{Rakshit_2020ApJS..249...17R}. First, we scale the AGN monochromatic luminosity (L$_{\rm 5100}$) by multiplying by a constant bolometric correction factor (=9.26) to get the bolometric luminosity for the source \citep{Richards_2006ApJS..166..470R}. While to derive the black hole mass, we adopt the following scaling relation from \cite{Vestergaard_2006ApJ...641..689V}:

\begin{equation}
\begin{split}
\log\left(\frac{M_{\rm BH}}{M_\odot}\right)
&= 0.91 + 0.5\,\log\left(\frac{L_{5100}}{10^{44}\,\mathrm{erg\,s^{-1}}}\right) \\
&\quad + 2\,\log\left(\frac{\mathrm{FWHM}}{\mathrm{km\,s^{-1}}}\right)
\end{split}
\end{equation}

Here, the relation assumes the FWHM to be that of the broad component of the H$\beta$ profile. We, then, estimate the Eddington ratio ($\lambda_{\rm Edd}$) by taking the ratio of the bolometric luminosity to the Eddington luminosity, i.e., $\approx$ 1.26 $\times 10^{38} \left(\frac{\rm M_{BH}}{\rm M_{\odot}}\right)$. The uncertainties for these derived values are estimated by propagating the uncertainties for the monochromatic luminosity and the FWHM(H$\beta_{\rm broad}$) as noted in Table \ref{tab:spec-table1}. We determine a black hole mass of log M$_{\rm BH}$ $\approx$ 7.107$\pm$0.053 (SDSS) and 7.157$\pm$0.050 (DESI), in units of solar masses. Correspondingly, the Eddington ratios were estimated to be (in log-scale) -2.108$\pm$0.054 (SDSS) and -1.769$\pm$0.052 (DESI), respectively.

\subsubsection{STEP2024phe/ZTF19aaflhnr}

\begin{figure*}[tp]
    \centering
    \includegraphics[width=\linewidth]{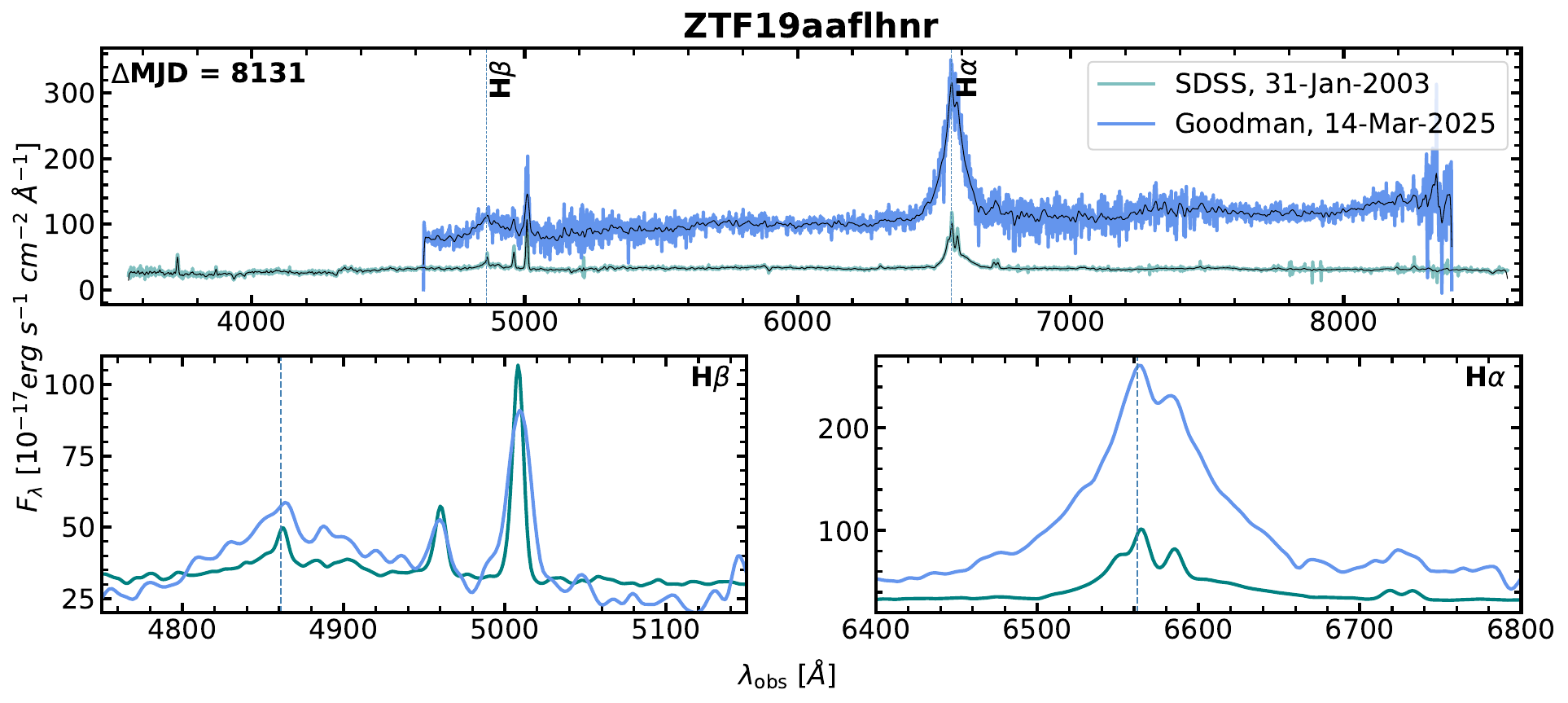}
    \caption{Alt Name: \textbf{2MASS J10511539+0548248}, Similar to Figure \ref{fig:spec_ztf18}, in lower panels, the SOAR/Goodman spectrum is shifted to match the continuum level for the SDSS (older) spectrum - to allow comparison of the change in the respective line profiles.}
    \label{fig:spec_ztf19}
\end{figure*}

\begin{figure*}[tp]
    \centering
    \includegraphics[width=\linewidth]{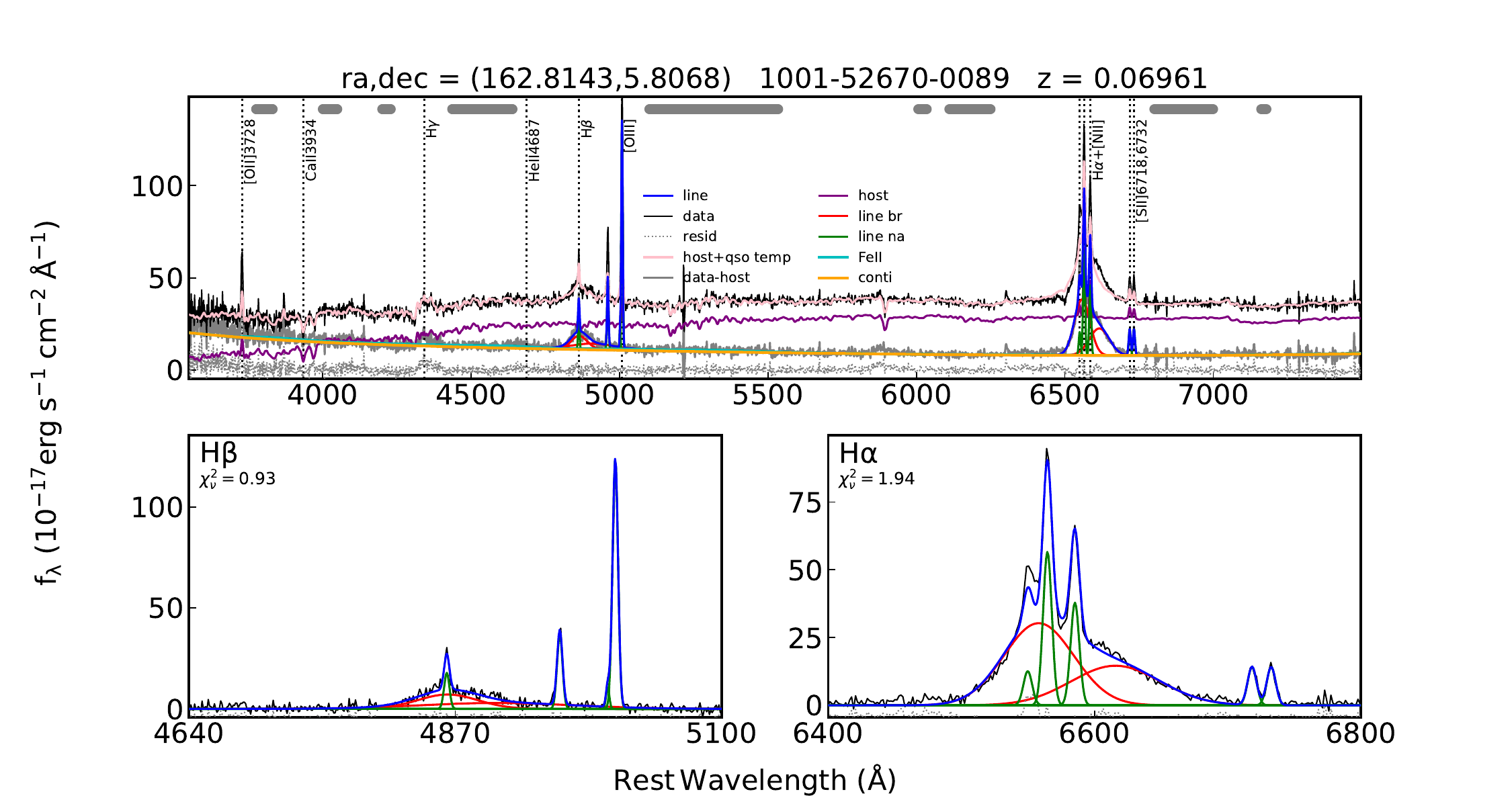}
    \caption{Similar to Figure \ref{fig:spec_sdss_ztf18} but for the source ZTF19aaflhnr's SDSS spectrum.}
    \label{fig:spec_sdss_ztf19}
\end{figure*}

\begin{figure*}[tp]
    \centering
    \includegraphics[width=\linewidth]{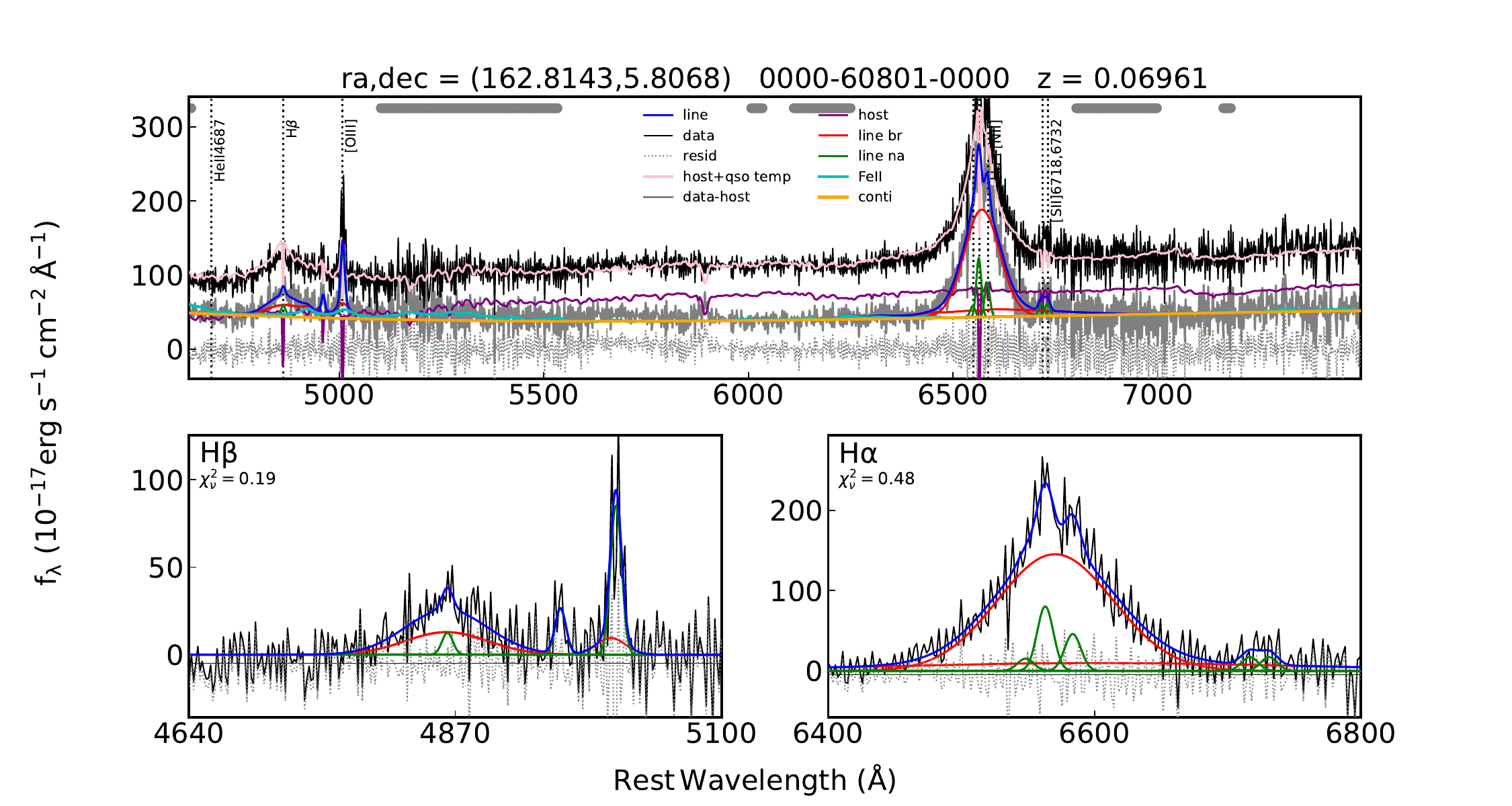}
    \caption{Similar to Figure \ref{fig:spec_sdss_ztf19} but for the more recent spectrum obtained using SOAR/Goodman.}
    \label{fig:spec_soar_ztf19}
\end{figure*}

For ZTF19aaflhnr target, we retrieved an archival SDSS spectrum (date: 31-January-2003; MJD 52670), and we managed to get a SOAR/Goodman spectrum from a dedicated observation (SO2024A-022; PI: De Bom \& Kilpatrick), which was observed more than 22 years after the SDSS epoch. As we can see from Figure \ref{fig:spec_ztf19}, the recent epoch (Goodman) has a higher total (and AGN) flux relative to the older, SDSS spectrum. The AGN flux for the more recent epoch is log L$_{\rm 5100}$ = 43.386$\pm$0.007 erg s$^{-1}$, which is $\sim$4 times higher than the older epoch (42.8$\pm$0.009 erg s$^{-1}$). The impact of the increased continuum flux can be clearly seen in the increased contributions of the emission lines' fluxes (see bottom panels in Figure \ref{fig:spec_ztf19}). Contrary to the previous target, this is a clear case of AGN brightening leading to the increased contribution to the total spectral output. While the host fraction reduced from 69.2\% to 53.9\% from the SDSS to SOAR/Goodman epochs, we note, however, that the SDSS spectrum demonstrates a notable redward asymmetry in both Balmer lines, which is a known characteristic for an AGN in the low flux state \citep{Bon_2012ApJ...759..118B}. 

In this case, we determine a black hole mass of log M$_{\rm BH}$ $\approx$ 7.648$\pm$0.062 (SDSS) and 8.025$\pm$0.044 (SOAR/Goodman), in units of solar masses. Correspondingly, the Eddington ratios were estimated to be (in log-scale) -1.982$\pm$0.062 (SDSS) and -1.773$\pm$0.044 (SOAR/Goodman), respectively. 

\begin{table*}[htp]
\centering
\caption{Basic and derived spectroscopic properties for the AGNs - I}
\label{tab:spec-table1}
\resizebox{\textwidth}{!}{%
\begin{tabular}{lcc|ccccr}
\hline
\textbf{Name} &
  \textbf{Facility} &
  \textbf{MJD} &
  \textbf{\begin{tabular}[c]{@{}c@{}}log L$_{\rm 5100}$\\ (erg s$^{-1}$)\end{tabular}} &
  \textbf{\begin{tabular}[c]{@{}c@{}}FWHM(H$\beta_{\rm br}$)\\ (km s$^{-1}$)\end{tabular}} &
  \textbf{\begin{tabular}[c]{@{}c@{}}log M$_{\rm BH}$\\ (M$_{\odot}$)\end{tabular}} &
  \textbf{log $\lambda_{\rm Edd}$} &
  \textbf{f$_{\rm host, 5100\,\mbox{\AA}}$} \\ \hline
ZTF18acvgziq & SDSS         & 53117 & 42.132$\pm$0.012 & 4665$\pm$223 & 7.107$\pm$0.053 & -2.108$\pm$0.054 & 88.6\% \\
             & DESI         & 59549 & 42.521$\pm$0.017 & 3112$\pm$178 & 7.157$\pm$0.050 & -1.769$\pm$0.052 & 69.4\% \\ \hline
ZTF19aaflhnr & SDSS         & 52670 & 42.8$\pm$0.009   & 4666$\pm$332 & 7.648$\pm$0.062 & -1.982$\pm$0.062 & 69.2\% \\
             & SOAR/Goodman & 60801 & 43.386$\pm$0.007 & 5141$\pm$259 & 8.025$\pm$0.044 & -1.773$\pm$0.044 & 53.9\% \\ \hline
\end{tabular}%
}
\end{table*}

\begin{table*}[htp]
\centering
\caption{Basic and derived spectroscopic properties for the AGNs - II}
\label{tab:spec-table2}
\resizebox{\textwidth}{!}{%
\begin{tabular}{lcc|cccccccr}
\hline
\textbf{Name} &
  \textbf{Facility} &
  \textbf{MJD} &
  \textbf{\begin{tabular}[c]{@{}c@{}}EW(H$\beta_{\rm br}$)\\ (\AA)\end{tabular}} &
  \textbf{\begin{tabular}[c]{@{}c@{}}FWHM([O{\sc iii}]$\lambda$5007)\\ (km s$^{-1}$)\end{tabular}} &
  \textbf{\begin{tabular}[c]{@{}c@{}}EW([O{\sc iii}]$\lambda$5007)\\ (\AA)\end{tabular}} &
  \textbf{\begin{tabular}[c]{@{}c@{}}FWHM(H$\alpha_{\rm br}$)\\ (km s$^{-1}$)\end{tabular}} &
  \textbf{\begin{tabular}[c]{@{}c@{}}EW(H$\alpha_{\rm br}$)\\ (\AA)\end{tabular}} &
  \textbf{\begin{tabular}[c]{@{}c@{}}FWHM([N{\sc ii}]$\lambda$6585)\\ (km s$^{-1}$)\end{tabular}} &
  \textbf{\begin{tabular}[c]{@{}c@{}}EW([N{\sc ii}]$\lambda$6585)\\ (\AA)\end{tabular}} &
  \textbf{\begin{tabular}[c]{@{}r@{}}EW([S{\sc ii}]$\lambda$6732)\\ (\AA)\end{tabular}} \\ \hline
ZTF18acvgziq & SDSS         & 53117 & 218.8 & 286.6 & 142.3 & 3070 & 273.7 & 353.2 & 94.7 & 31.3 \\
             & DESI         & 59549 & 73.9  & 583.1 & 48.5  & 2823 & 190.9 & 456.8 & 27.3 & 10.8 \\ \hline
ZTF19aaflhnr & SDSS         & 52670 & 86.1  & 341.9 & 67.9  & 4032 & 403.3 & 353.2 & 39.1 & 15.0 \\
             & SOAR/Goodman & 60801 & 51.9  & 680.2 & 24.5  & 4856 & 452.6 & 658.8 & 16.2 & 6.0  \\ \hline
\end{tabular}%
}
{\scriptsize NOTES: FWHMs for the [N{\sc ii}] and the [S{\sc ii}] doublets are individually tied to the same value. The associated uncertainties are not reported for brevity, but are concordant with the values reported in Table \ref{tab:spec-table1}. The doublets are modeled assuming a proportion of 1:3.}
\end{table*}

\subsubsection{Light Curve Analysis - AGN variability}

When searching for electromagnetic counterparts to BBH mergers embedded in AGN disks, several factors can limit the detectability and interpretation of transient candidates. One of the main challenges is distinguishing a merger-induced flare from the intrinsic optical variability of the AGN itself. These two phenomena can appear similar in single-epoch observations, but they are expected to differ statistically in duration, amplitude relative to the baseline flux, and temporal behavior.

AGN variability is typically stochastic and can be characterized using pre-event light curves, whereas a merger-related flare is expected to produce a more localized excess in flux above the historical baseline. Therefore, comparing the candidate light curve with its long-term variability history prior to the GW trigger, and performing extended post-event monitoring, provides important diagnostic power for assessing the origin of the flare.

\begin{figure*}[tp]
    \centering
    \includegraphics[width=1\linewidth]{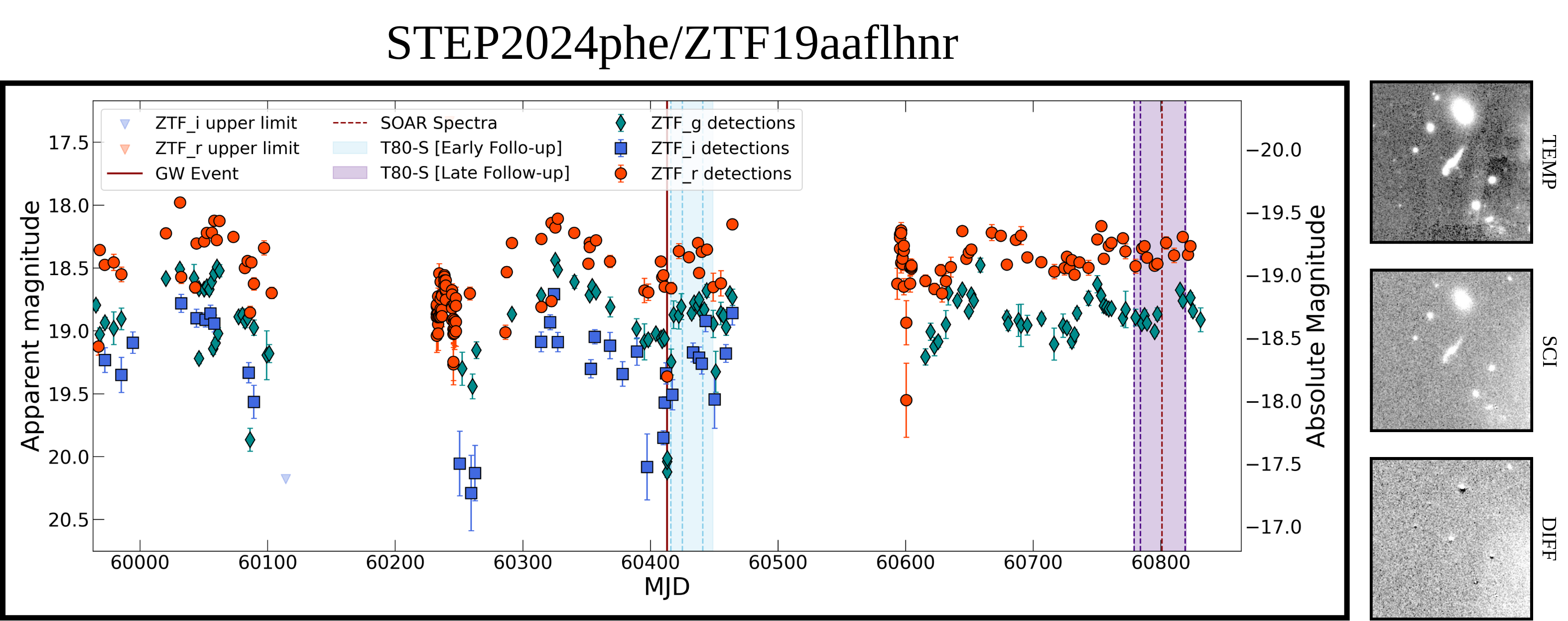}
    \caption{Light curve of ZTF19aaflhnr (STEP2024phe).}
    \label{fig:candidates_ztfdr23_lc}
\end{figure*}

\begin{figure*}[tp]
    \centering
    \includegraphics[width=1\linewidth]{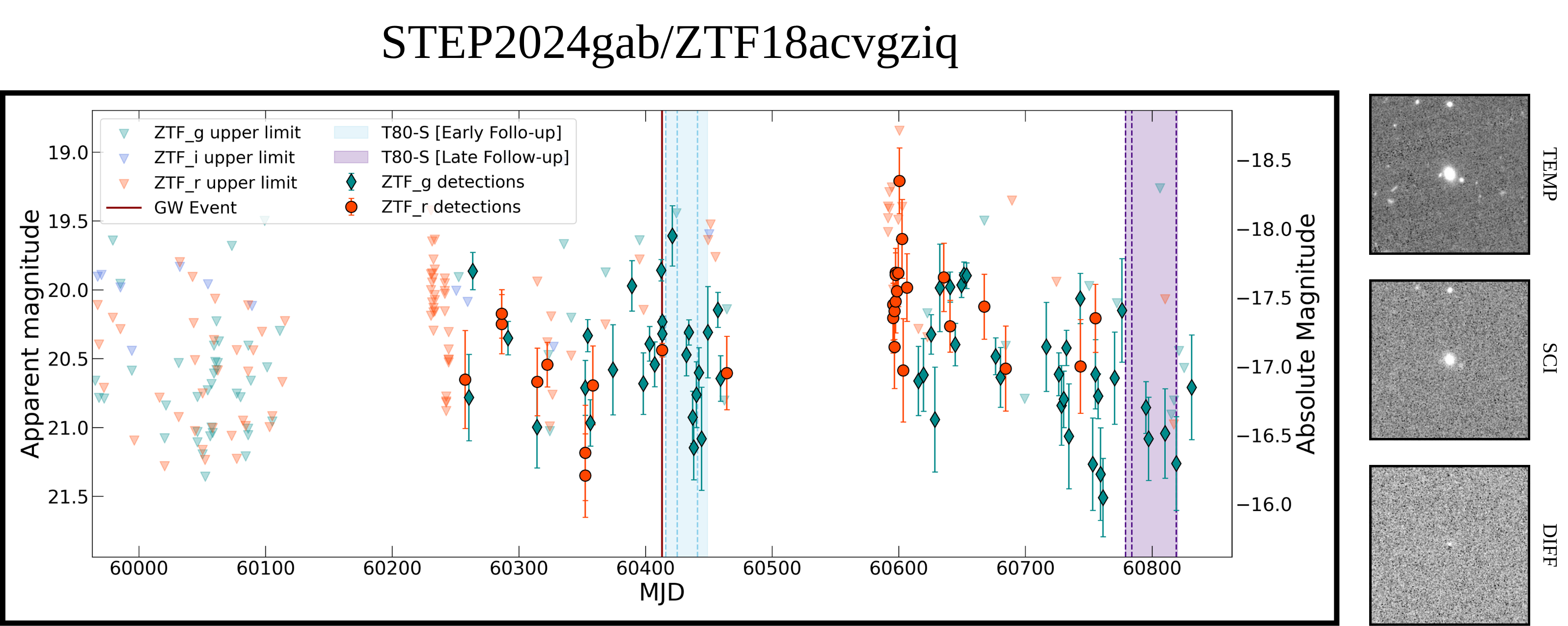}
    \caption{Light curve of ZTF18acvgziq (STEP2024gab).}
    \label{fig:lc_gab}
\end{figure*}

Because the predicted delay time between the GW merger and any associated electromagnetic flare depends on model assumptions and on the location of the merger within the AGN disk, continued temporal coverage is essential to search for delayed brightening consistent with dark-flare scenarios. Motivated by this, we retrieved ZTF forced-photometry measurements from difference imaging spanning a 600-day window centered on the merger date. We excluded all measurements with signal-to-noise ratio $\mathrm{S/N} < 3$, following the recommendations of the ZTF forced-photometry reference, and applied the standard forced-photometry quality cuts described in Section~6.1 (Quality Filtering) of \citet{ZTF_forced_photometry_ref}.
Figures~\ref{fig:candidates_ztfdr23_lc} and \ref{fig:lc_gab} show the light curves of ZTF19aaflhnr and ZTF18acvgziq, respectively. The vertical lines indicate the time of the GW merger and the epochs at which the follow-up observations were obtained. Inspection of the light curves reveals no clear evidence of a transient temporally associated with the GW event. The detections above the baseline flux are most likely explained by stochastic AGN variability, which commonly produces irregular optical fluctuations on comparable timescales. 

If a compact-object merger occurred within the accretion disk of either AGN, several scenarios could still result in the absence of a detectable electromagnetic counterpart. First, any merger-driven flare could be outshone by the intrinsic AGN emission, preventing its detection in difference imaging. Second, the remnant black hole may not have remained embedded in the disk long enough to accrete sufficient material or launch relativistic jets capable of producing a luminous flare. In addition, some models predict that if the merger remnant receives a recoil kick approximately perpendicular to the disk plane, it may rapidly exit the disk through the low-density cavity region, strongly suppressing post-merger accretion and the resulting electromagnetic emission \cite{Kimura2021,RodrguezRamrez2025}.

Even when a BBH merger successfully launches outflows (or jets depending on the emission scenario), the detectability of the resulting breakout emission depends strongly on the AGN viewing geometry. In Type~II (usually observed edge-on) systems, the outflow propagates perpendicular to the line of sight, rendering the breakout emission unobservable. Type~I (approximately face-on) AGNs are geometrically more favorable, as the outflow is expected to emerge on the observer-facing side of the disk roughly half of the time. The AGN inclination therefore acts as an additional selection effect that modulates the observable fraction of BBH-induced flares, independent of the intrinsic emission properties of the merger \citep{McfactsIV_Mcpike}.


\subsection{Optical Counterparts to Binary Black Hole Mergers}\label{sec:bbh_optical}

Having ruled out all known optical transients as viable counterparts to S240413p, we further investigate the expected properties of dark flares within the context of our search strategy by adopting the emission mechanism described in \cite{RodrguezRamrez2025}. In this scenario, similar to that proposed by \cite{Kimura2021}, the binary merger occurs inside an underdense cavity created by the binary in the thin accretion disk prior to coalescence. The dark flare is produced when the recoiling remnant black hole exits this cavity and encounters denser regions of the disk, triggering hyper-Eddington accretion. 

Under highly super-Eddington inflow conditions, the remnant is expected to drive powerful radiation-driven winds and potentially launch bipolar collimated jets. In either case, the model assumes that the remnant captures disk material while simultaneously driving a wide-angle quasi-spherical outflow as it propagates through the AGN disk. We focus here on the thermal emission produced as this outflow-driven ejecta expands above and below the disk plane; as the material cools, the photosphere recedes and optical and ultraviolet photons escape, giving rise to the observable transient.

This model takes into account the time required for the remnant to exit the underdensity cavity and enter the unperturbed disk ($\Delta t_{\rm cav}$), the time needed to accumulate sufficient material to initiate the wind ($\Delta t_{\rm hl}$), the time required for the wind to reach the upper and lower boundaries of the disk ($\Delta t_{\rm bo}$), and the photon diffusion time until radiation can escape the ejecta ($\Delta t_{\rm phot}$). The total delay between the GW event and the electromagnetic emission is therefore approximately given by

\[
\Delta t_{\rm delay} \approx (1+z)\left(\Delta t_{\rm cav} + \Delta t_{\rm hl} + \Delta t_{\rm bo} + \Delta t_{\rm phot}\right),
\]
with $z$, being the redshift of the source.
The duration of the flare is given by Equation~36 of \citet{RodrguezRamrez2025}:

\[
\Delta t_{\text{duration}} = (1 + z) \sqrt{ \frac{9 \kappa_{\text{T}} M_0}{2 \pi^3 c u_0} } ,
\]

where

\[
u_0 = \left( \frac{E_0}{M_0} \right)^{1/2}.
\]

Here $M_0$ represents the mass
expelled by the remnant driven wind,
$E_0$ is the energy stored by the BH wind in each ejection (see Equation~15 of \citet{RodrguezRamrez2025}), $\kappa_{\rm T}$ is the Thomson electron-scattering opacity, and $c$ is the speed of light. In summary, the observable flare properties depend on both the AGN host parameters and the merger characteristics. The relevant host parameters are the SMBH mass ($M_{\rm SMBH}$), the accretion rate ($\lambda_{\rm Edd}$) in Eddington units, and the disk viscosity parameter $\alpha_d$. The merger parameters include the remnant mass $M_{\rm rem}$, the recoil velocity $v_\mathrm{k}$, the kick angle $\theta_\mathrm{k}$ (relative to the disk plane), the merger location within the disk $R_{BH}$, and the efficiencies of remnant accretion ($\eta_a$) and wind production ($\eta_w$). A more detailed description of the physical model and its assumptions can be found in \cite{RodrguezRamrez2025}.

\subsubsection{Constraints on EM Counterparts to S240413p}

Following \citet{Kilpatrick_2021_gravcole} and \citet{gravicoulter2024gravitycollectivecomprehensiveanalysis}, and the BBH extension of \citet{Darc25_multimomel_constraint_bbh}, we assess which BBH merger configurations would yield detectable flares under our observing strategy by comparing the limiting magnitudes, observation times, and filters of our imaging data directly against model light curves.  While \citet{Darc25_multimomel_constraint_bbh} derived constraints across multiple BBH flare models \citep{McKernan2019, tagawa_phe_model, juan_publicado}, here we focus exclusively on the radiation-driven outflow scenario of \citet{RodrguezRamrez2025}, hereafter JRR+25, which remains largely unexplored observationally.

We adopt a $3\sigma$ detection threshold to define our limiting magnitudes throughout. To quantify detection efficiency, we use \texttt{Teglon} \citep{tegloncoulter_2021_5683508}, an open-source tool for gravitational-wave follow-up analysis. \texttt{Teglon} ingests the telescope pointing information (R.A., Dec., filter, time, and $3\sigma$ limiting magnitude), cross-matches each pointing with the corresponding HEALPix pixels, and retrieves the GW-derived luminosity distance distribution for each pixel. It then re-expresses the limiting magnitude as the maximum distance at which a modeled EM source of given time-dependent luminosity and line-of-sight extinction would be detectable. A full description of the \texttt{Teglon} detection efficiency formalism is provided in Appendix~A of \citet{gravicoulter2024gravitycollectivecomprehensiveanalysis}.

The resulting detection efficiency is interpreted as the probability that a simulated light curve from JRR+25 would have been detected by our observational campaign. The maximum achievable detection probability is fundamentally limited by the total sky-localization probability covered by our dataset. Since the model is defined over five parameters ($v_\mathrm{kick}$, $M_\mathrm{rem}$, $M_\mathrm{SMBH}$, $R_\mathrm{BH}$, $\theta_\mathrm{kick}$), the detection efficiency spans a five-dimensional parameter space. We present results through two complementary two-dimensional projections.

The first is the \textbf{mean-probability map}, in which the detection efficiency for a given parameter pair is averaged over all remaining dimensions. This marginalization reveals systematic trends across parameter space and identifies regions that are more likely to produce detectable flares on average. The second is the \textbf{maximum-probability map}, in which we record the highest detection efficiency obtained over all remaining parameter combinations for each pair. These maps identify regions where detectable flares are possible under at least one favorable configuration; conversely, regions of uniformly low maximum probability indicate configurations unlikely to produce detectable emission regardless of the remaining parameter values.

\textbf{Prior choice:} To generate the light curve simulations for a BBH merger, we adopt prior distributions motivated by \citet{Darc25_multimomel_constraint_bbh} and \citet{RodrguezRamrez2025} for the JRR+25 scenario. For the remnant black hole's recoil (kick) velocity following the merger, we consider values in the range 100--500~km\,s$^{-1}$, guided by the velocity distribution predicted by the JRR+25 model. In addition, the post-merger kick angle must be sufficiently small to allow the remnant to interact with the denser regions of the AGN disk. We therefore explore three representative kick angles: 1$^\circ$, 5$^\circ$, and 15$^\circ$.

The distance from the supermassive black hole is constrained by the \citet{srikogoodman10.1046/j.1365-8711.2003.06431.x} disk model, where we explore a range from 500 $R_g$ up to $R_{\rm max}$, with $R_{\rm max}$ proposed as (see \cite{RodrguezRamrez2025} for details):
\begin{equation}
    R_{\rm max} = 10^4 \left(\frac{M_{\rm SMBH}}{10^8\,M_\odot}\right)^{-1/2} R_g.
\end{equation}

For the remnant black hole mass, we adopted a range of $20$--$160~M_\odot$, which encompasses the majority of BBH mergers observed by the LVK collaboration, with the notable exception of GW231123, reported to have a total mass of $\sim 237~M_\odot$. To maintain an analysis that is as agnostic as possible with respect to the observer's viewing direction, we randomly assigned the side from which the bubble eruption occurs in each simulation. Given the computational cost of generating a complete set of simulations for each SMBH mass--approximately 10 hours per mass value across the full adopted parameter grid--our analysis focuses on a set of representative masses: $10^6$, $10^7$, $10^8$, and $10^9~M_\odot$, with a fixed accretion rate of 0.05.

\begin{table}[h]
\centering
\caption{Discrete values of distance of remnant black hole from the SMBH \(R_{\mathrm{BH}}\) adopted for each supermassive black hole mass \(M_{\mathrm{SMBH}}\). Radii are given in units of gravitational radius \(R_g = GM_{\mathrm{SMBH}}/c^2\). }
\label{tab:Rbh_grid}
\begin{tabular}{cc}
\hline
\hline
$M_{\mathrm{SMBH}}\;[M_\odot]$ & $R_{\mathrm{BH}}\;[R_g]$ \\
\hline
$10^{9}$ & $500 - 3000$ \\
$10^{8}$ & $500 - 10000$ \\
$10^{7}$ & $500 - 30000$ \\
$10^{6}$ & $500 - 30000$ \\
\hline
\end{tabular}
\end{table}

\begin{table*}[htp]
\centering
\caption{Adopted Prior Distributions for JRR+25 BBH Merger Simulations}
\label{tab:priors_jrr25}
\begin{tabular}{lccc}
\hline
Parameter & Symbol & Range / Values & Units \\
\hline
Remnant black hole mass & $M_\mathrm{bh}$& 20 -- 160 & $M_\odot$ \\
SMBH mass & $M_\mathrm{SMBH}$ & $10^6$, $10^7$, $10^8$, $10^9$ & $M_\odot$ \\
Remnant kick velocity & $v_\mathrm{kick}$ & 100 -- 500 & km\,s$^{-1}$ \\
Remnant kick angle & $\theta_\mathrm{kick}$ & 1, 5, 15 & degrees \\
\hline
\end{tabular}
\end{table*}

In addition, we also simulated optical light curves using the JRR+25 model, adopting a more realistic prior motivated by the possibility that a detectable flare could arise from a BBH merger occurring within the AGN disks of the candidate hosts ZTF19aaflhnr and ZTF18acvgziq. The SMBH mass and accretion rate were fixed to the observational values derived from spectral fitting. Specifically, we adopt the parameters inferred from the Goodman spectrum of ZTF19aaflhnr (MJD 60801) and the DESI spectrum of ZTF18acvgziq (MJD 59549) as fiducial inputs.

For this realistic prior we adopt conservative values for the efficiency parameters, $\eta_a=\eta_{\rm w}=0.05$, and sample the remaining model parameters within the ranges $M_{\rm rem}\in[10,150]\,M_\odot$, $v_k\in[100,500]\,{\rm km\,s^{-1}}$, $\theta_k\in[-90^\circ,90^\circ]$, and $a/R_g\in[100,R_{\rm max}/R_g]$, where $R_g\equiv GM_{\rm BH}/c^2$. The kick velocity and direction ($v_k$, $\theta_k$) are sampled from uniform distributions, while the merger location $a/R_g$ is sampled uniformly in logarithmic space. The remnant mass is drawn from a power-law distribution $\propto M^{-\alpha_m}$ with $\alpha_m=3.5$, motivated by the inferred distribution of total black hole masses in GWTC-3 \cite{Abbott_2023}. Although the GWTC-3 mass distribution exhibits additional structure beyond a simple power law, this approximation provides a more realistic description of the remnant mass distribution than assuming a uniform prior.

\section{Results \& Discussion }

\begin{figure*}[tp]
    \centering
    \begin{subfigure}{0.8\textwidth}
        \centering
        \includegraphics[width=\textwidth]{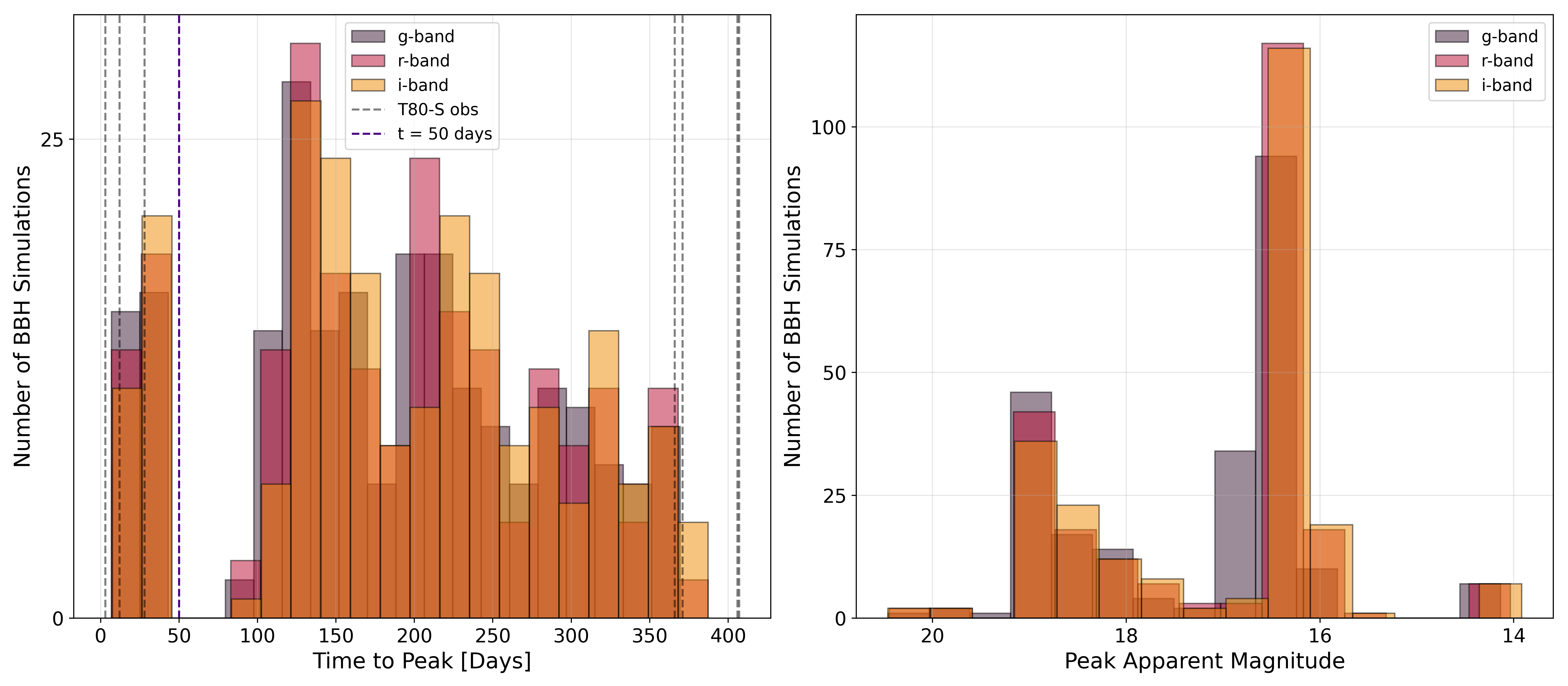}
        \caption{Simulated BBH flare properties with detection probabilities above 50\%. Left: peak magnitude distribution for each band (g, r, i). Right: time to peak after the GW event. Vertical dashed gray lines indicate follow-up observation epochs.}
        \label{fig:fig1a}
    \end{subfigure}
    
    \vspace{0.5cm} 
    
    \begin{subfigure}{0.8\textwidth}
        \centering
        \includegraphics[width=\textwidth]{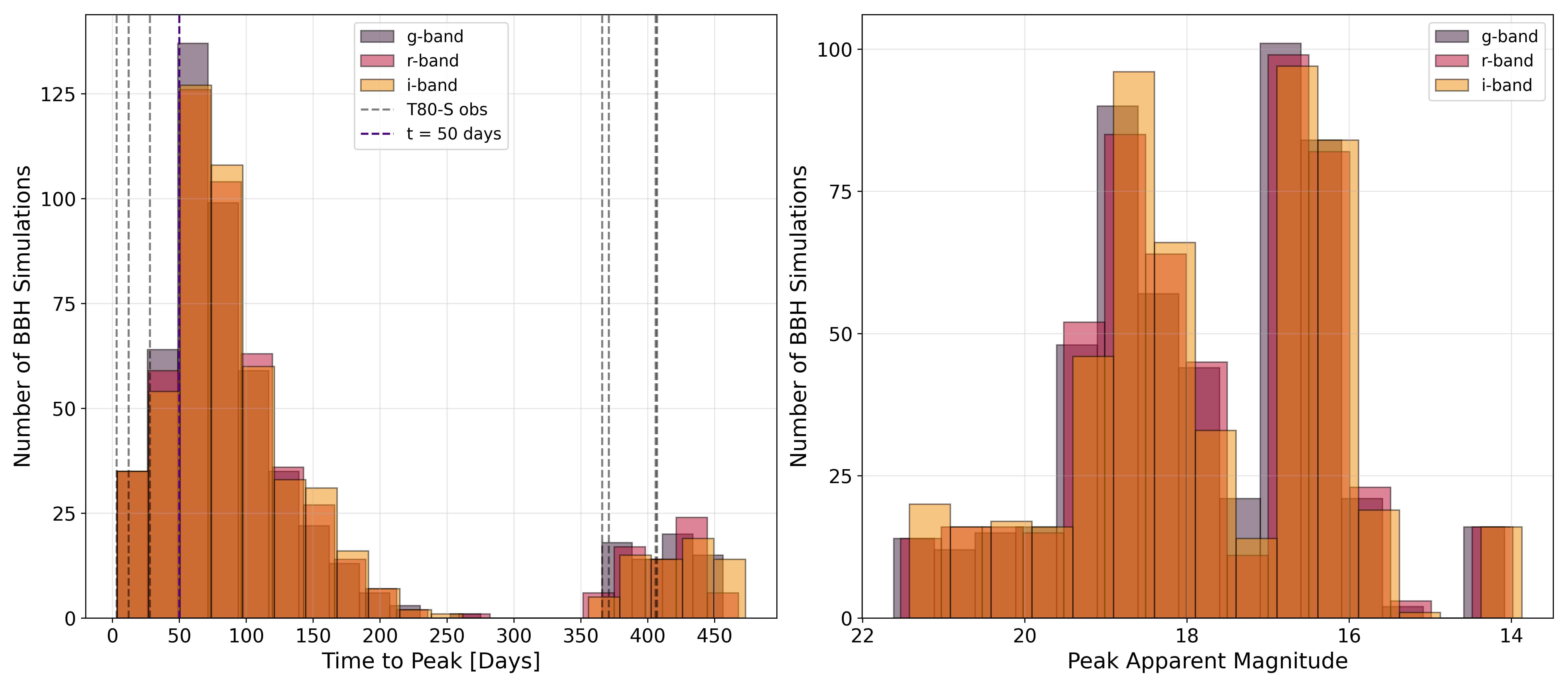}
        \caption{Simulated BBH flare properties with detection probabilities below 50\%. Left: peak magnitude distribution for each band (g, r, i) for flares not detected or with probability below 50\%. Right: time to peak after the GW event. Vertical dashed gray lines indicate follow-up observation epochs.}
        \label{fig:fig1b}
    \end{subfigure}

    \caption{BBH flare properties from the JRR+25 model. Panel (a) shows detectable flares (probability $>$50\%), panel (b) shows low-probability flares (probability $<$50\%). Each panel presents both the peak magnitude distributions and time-to-peak distributions for the g, r, and i bands.}
    \label{fig:merged_flares}
\end{figure*}

\begin{figure*}[tp]
    \centering
    \includegraphics[width=1\linewidth]{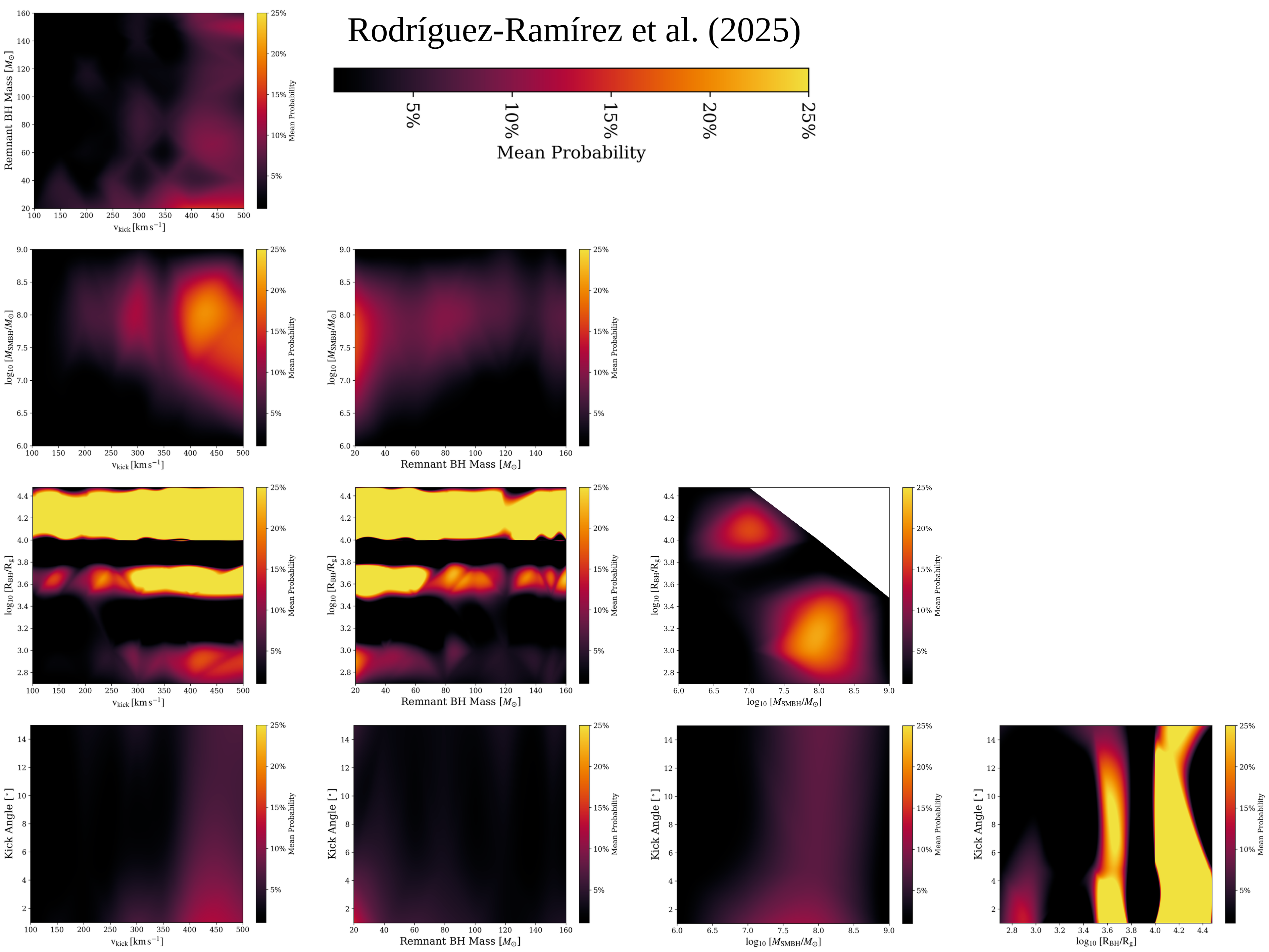}
    \caption{Corner plot showing the two-dimensional projections of the mean detection probability predicted by the JRR+25 model as a function of the kick velocity ($v_{\mathrm{kick}}$), remnant black hole mass ($M_{\mathrm{BH}}$), SMBH mass ($\log_{10}(M_{\mathrm{SMBH}}/M_\odot)$), distance from the AGN ($\log_{10}(R_{\mathrm{BH}}/R_{\mathrm{g}})$), and kick angle ($\theta_{\mathrm{kick}}$). Each individual 2D-plot represents the mean detection efficiency averaged over all other parameters not shown in the corresponding projection. Darker colors indicate low mean detection probabilities, while brighter yellow regions correspond to higher probabilities, highlighting favorable configurations for producing observable electromagnetic flares associated with BBH mergers in AGN disks.}

    \label{fig:jrr25_mean}
\end{figure*}

\begin{figure*}[tp]
    \centering
    \includegraphics[width=1\linewidth]{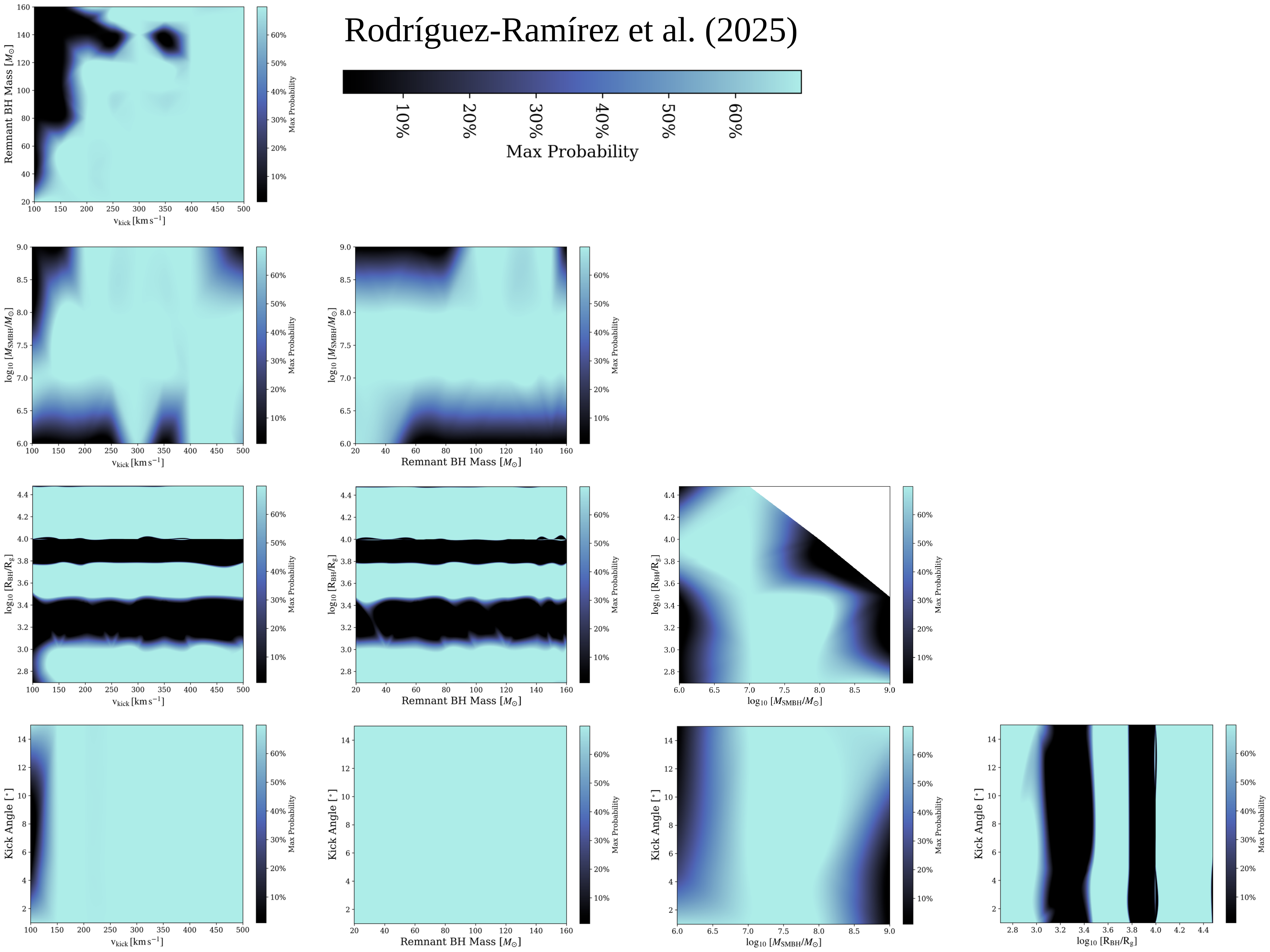}
    \caption{Corner plot showing the two-dimensional projections of the maximum detection probability as a function of the kick velocity ($v_{\mathrm{kick}}$), remnant black hole mass ($M_{\mathrm{BH}}$), SMBH mass ($\log_{10}(M_{\mathrm{SMBH}}/M_\odot)$), distance from the SMBH ($\log_{10}(R_{\mathrm{BH}}/R_{\mathrm{g}})$), and kick angle ($\theta_{\mathrm{kick}}$). Each panel shows the maximum detection probability obtained over all other parameters not displayed in the corresponding projection, highlighting whether at least one model realization within that region of parameter space can produce a detectable electromagnetic counterpart. Black regions indicate that none of the simulations produced a detectable EM counterpart under our observing strategy, whereas light blue regions identify parameter combinations for which at least one model predicts a detectable dark flare.}

    \label{fig:jrr25_max}
\end{figure*}

In this work, we employ the JRR+25 light-curve simulations over a broad prior distribution and compare them with our observational limits using the \texttt{Teglon} software. This approach allows us not only to assess how such flares can be detected under our follow-up strategy, but also to identify the regions of parameter space and AGN environments in which detectable emission are most likely to occur given our observational constraints.

Figure~\ref{fig:merged_flares} illustrates the properties of the simulated JRR+25 flares separated by detection probability. Panel~(a) shows flares with detection probabilities above 50\%, which peak at magnitudes accessible to our imaging depth across the $g$, $r$, and $i$ bands and concentrate at time-to-peak values consistent with our follow-up epochs. Panel~(b) shows the complementary population of low-probability flares, which are either too faint to reach our limiting magnitudes or peak outside our observing windows, as indicated by the vertical dashed lines marking our observation epochs. As described in Section~\ref{sec:methods}, we present two-dimensional projections of the multidimensional parameter space for all combinations of parameter pairs in a corner-plot style representation. Figure \ref{fig:jrr25_mean} shows the mean detection probability marginalized over all remaining parameters ($M_\mathrm{SMBH}$, $M_\mathrm{bh}$, $R_\mathrm{bh}$, $\theta_\mathrm{kick}$, and $v_\mathrm{kick}$) \footnote{This is not a posterior distribution. The complete detection-efficiency surface is defined in a five-dimensional parameter space.}. Figure \ref{fig:jrr25_max} presents the corresponding maximum detection probability for each parameter pair.

We first focus on the two most relevant parameters that can be
constrained from the gravitational-wave strain data: the remnant
black hole mass, approximated as the sum of the component masses,
and the recoil velocity. The kick velocity arises from anisotropic
gravitational-wave emission during the merger and can be related,
through the empirical formulae of \citet{Lousto_2010,Lousto_2012},
to the binary mass ratio ($q = M_1/M_2$) and the effective spin
parameter ($\chi_{\mathrm{eff}}$). The two-dimensional projections of the detection-efficiency maps reveal a clear trend toward higher detectability for mergers with large recoil velocities ($v_\mathrm{kick} \gtrsim 350$--$500~\mathrm{km\,s^{-1}}$), visible across all panels sharing the $v_\mathrm{kick}$ axis in the mean-probability maps (Fig.~\ref{fig:jrr25_mean}). The maximum-probability maps (Fig.~\ref{fig:jrr25_max}) show that at least one realization yields detectable emission across nearly the full kick-velocity range, though such configurations are not representative of the average behavior.

In the regime of low remnant masses ($M_\mathrm{rem} < 40~M_\odot$) and high kick velocities ($v_\mathrm{kick} > 350~\mathrm{km\,s^{-1}}$), the predicted flare properties depend sensitively on the SMBH mass. For $M_\mathrm{SMBH} = 10^8~M_\odot$, detectable flares peak between $\sim$100 and 250~days post-merger with prolonged emission lasting $\mathcal{O}(100~\mathrm{days})$, whereas for $M_\mathrm{SMBH} = 10^7~M_\odot$ they peak before $\sim$50~days with a sharper maximum and shorter duration. This behavior reflects the dependence of the disk scale height on SMBH mass: lower-mass AGN disks are geometrically thinner, allowing the remnant outflow to reach the disk edge more rapidly and producing earlier, shorter-lived flares. 

This trend is most pronounced at $M_\mathrm{SMBH} = 10^6~M_\odot$, where all detectable models peak within $\sim$30~days of the merger. Nevertheless, the $v_\mathrm{kick}$--$M_\mathrm{SMBH}$ maximum-probability map reveals a region of uniformly low detectability at $M_\mathrm{SMBH} = 10^6~M_\odot$ and $v_\mathrm{kick} \lesssim 250~\mathrm{km\,s^{-1}}$, arising from two main effects: many simulations in this regime fail to produce significant emission, and when emission does occur it peaks at intermediate delays of $\sim$70--150~days post-merger, falling largely outside our follow-up windows.

Another key parameter inferred from the GW strain data is the remnant black hole mass. The maximum-probability maps show no strong preference for any particular remnant mass; under favorable disk conditions and recoil configurations, detectable flares can be produced across a broad mass range. Our follow-up strategy is therefore, to first order, agnostic to the remnant mass of the GW event --- an advantageous property given that no precise mass information is available at the time of follow-up.

For very massive remnants ($M_\mathrm{rem} > 120~M_\odot$), JRR+25 simulations show that most detectable flares emerge only after $\sim$90~days post-merger. The exception occurs when massive remnants receive high recoil velocities ($v_\mathrm{kick} = 500~\mathrm{km\,s^{-1}}$) in AGN disks with $M_\mathrm{SMBH} = 10^8~M_\odot$, where detectable emission can arise at earlier times.

Investigating the AGN environments in which electromagnetic counterparts are most likely to be detectable may help inform the optimal observational strategy one must adopt when searching for dark flares. In this context, we focus on the mass of the central SMBH and the merger distance from the SMBH, parameterized as
$\log_{10}(R_\mathrm{bh}/r_g)$, where $r_g$ is the gravitational radius.

In the $R_\mathrm{rem}$--$M_\mathrm{SMBH}$ mean probability map (Figure \ref{fig:jrr25_mean}), two prominent high-probability regions emerge. The first is centered on AGNs with $M_\mathrm{SMBH} = 10^8~M_\odot$ and $\log_{10}(R_\mathrm{rem}/r_g) \approx 2.8$ -- 3.4. The second appears for $M_\mathrm{SMBH} = 10^7~M_\odot$ and $\log_{10}(R_\mathrm{rem}/r_g) \approx 3.8$ -- 4.4. Converting these distances using the definition of $r_g$, both regions correspond to physical merger distances of approximately 0.003--0.012~pc from the SMBH. Selecting configurations with detection probabilities above 50\%, we find that these regions are associated with long-duration flares and delayed peaks occurring $\gtrsim$ 100--250 days after the merger (Figure~\ref{fig:merged_flares}). Notably, this portion of parameter space is only accessible due to our late-time follow-up campaign. In contrast, the maximum-probability map shows that the two high-probability concentrations largely dissolve, with at least one detectable flare occurring across the full range of merger distances. Additionally, lower-mass AGNs ($M_\mathrm{SMBH} = 10^6~M_\odot$) favor mergers at larger radii, whereas more massive AGNs ($M_\mathrm{SMBH} = 10^9~M_\odot$) favor mergers closer to the central SMBH.

The $M_\mathrm{rem}$--$M_\mathrm{SMBH}$ projection reveals a complementary trend. Lower-mass remnants ($M_\mathrm{rem} < 40~M_\odot$), associated with comparatively lower accretion-powered luminosities in the JRR+25 model, produce detectable flares primarily in $M_\mathrm{SMBH} = 10^6$--$10^8~M_\odot$ environments. Higher-mass remnants ($M_\mathrm{rem} > 80~M_\odot$) begin to show detectable configurations in more massive AGNs, particularly in the $10^8$--$10^9~M_\odot$ regime.

The high-probability stripes visible in the $R_\mathrm{bh}$ projections are an artifact of the sparse sampling of our simulation grid; a finer and more uniformly sampled grid would smooth these transitions. More broadly, BBH-driven optical flare models remain relatively unexplored, and no single emission mechanism has been established to govern the full range of thermal and non-thermal processes that arise when a remnant black hole accretes at super-Eddington rates through AGN disk material.

In summary, AGN environments with $M_\mathrm{SMBH} \sim 10^7$--$10^8~M_\odot$ represent the most favorable sites for producing detectable counterparts in the JRR+25 scenario, as they balance a moderately thin disk geometry with sufficient material for super-Eddington accretion. In lower-mass AGNs ($M_\mathrm{SMBH} \sim 10^6~M_\odot$), the disk is geometrically thinner and most simulations indicate that the remnant does not spend enough time within the disk to accrete sufficient material for detectable emission. Conversely, in very massive AGNs ($M_\mathrm{SMBH} \sim 10^9~M_\odot$), the disk is significantly thicker, pushing predicted flare delays beyond $\sim$450~days --- timescales at which a reliable association between the electromagnetic transient and the GW event becomes increasingly difficult to establish.

\begin{figure*}[tp]
    \includegraphics[width=0.95\linewidth]{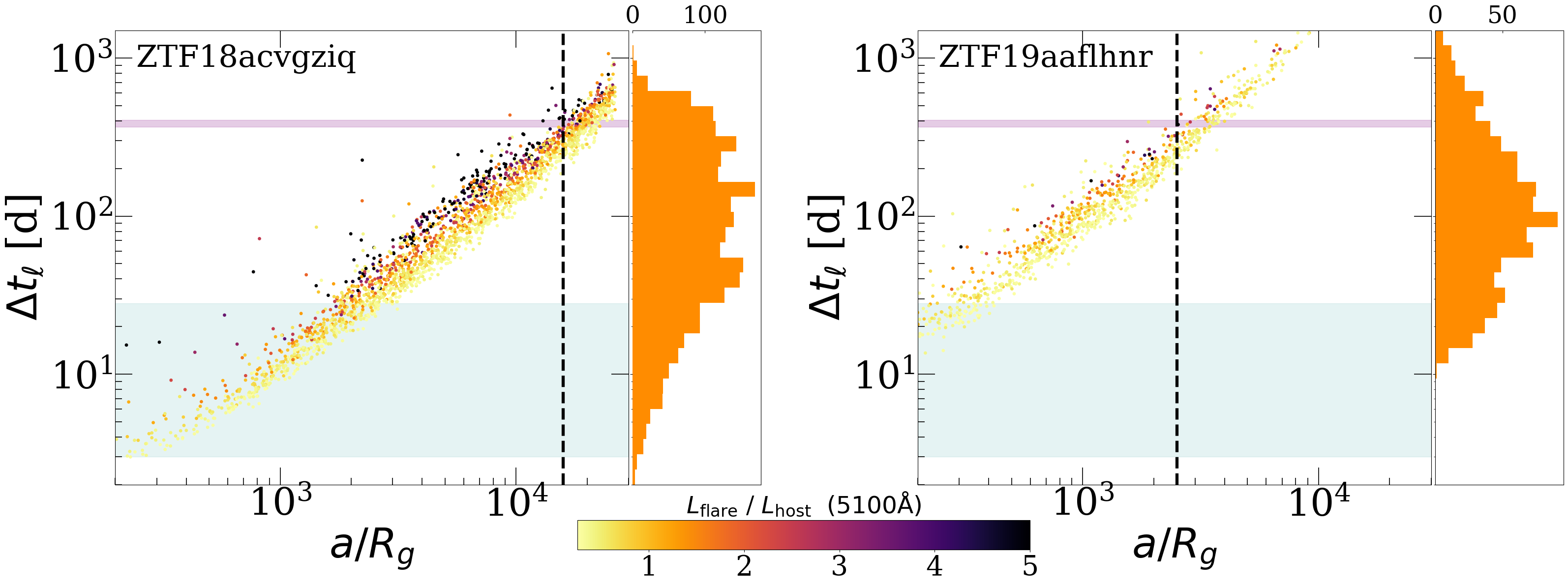}
    \caption{Distribution of the expected time delays of the EM counterpart predicted by the AGN disk merger scenario of \cite{RodrguezRamrez2025} for the sources ZTF18acvgziq and ZTF19aaflhnr. The simulated flares assume AGN hosts with the spectroscopic parameters reported in Table~\ref{tab:spec-table1}, while the remaining model parameters are sampled from the realistic prior described in Sec. \ref{sec:methods}. Colored points denote simulated flares with emission ratios (flare-to-host luminosity) $>0.25$ at 5100~\AA, as indicated by the color bar. The dashed vertical lines mark the locations of the thermal migration traps predicted by \cite{Grishin_2024} for each source \footnote{The radiation efficiency used in \cite{Grishin_2024} to define the Eddington accretion rate is $\eta=0.06$, whereas the framework of JRR+25 adopts $\eta=0.1$. This difference is taken into account when estimating the trap locations.}. The lower and upper shaded regions indicate the epochs of early and late follow-up observations, respectively, where the zero point on the $y$ axis corresponds to the merger time of S240413p.}
    \label{fig:At_dist}
\end{figure*}

\subsection{Expected Optical Flare Delays in the AGN Hosts ZTF18acvgziq and ZTF19aaflhnr}

We compare our observational follow-up windows with the delay times predicted for optical flares produced by BBH mergers embedded in the AGN disks of ZTF18acvgziq and ZTF19aaflhnr. To do so, we adopt the emission model developed by \citet{RodrguezRamrez2025} and use the spectroscopically derived host properties ($z$, $M_{\rm SMBH}$, $\lambda_{\rm Edd}$, and $L_{5100}$) reported in Table~\ref{tab:spec-table1}.

Figure~\ref{fig:At_dist} shows the distribution of predicted flare delays in the $\Delta t_\ell$--$R_\mathrm{BH}/R_{\rm g}$ plane for both hosts, computed using the realistic priors described in Section~\ref{sec:methods}. Each point corresponds to a simulated flare with intrinsic luminosity ratio $\nu L_{\nu}/L_{5100} > 0.25$.

TThe radial distribution of BBH mergers within AGN disks remains uncertain. Migration traps, locations where disk torques change sign and inward-migrating objects stall,  represent one of the most physically motivated mechanisms for identifying preferred merger sites. In Fig. \ref{fig:At_dist} we indicate with vertical dashed lines the locations of the thermal migration traps predicted by \citet{Grishin_2024}. For the disk parameters of ZTF18acvgziq and ZTF19aaflhnr, these occur at $a/R_{\rm g}\approx10^{4.2}$ and $10^{3.4}$, respectively. The predicted dark flare delays span from tens to several hundred days, reflecting the wide range of possible merger radii. The early follow-up window (cyan shaded region) overlaps only with the tail of the delay distribution for both sources, making the detection of an early flare unlikely. The late follow-up epoch (purple shaded region) approximately coincides with the peak of the delay distribution and the migration trap locations in both hosts, suggesting that if a BBH-induced flare occurred in either AGN, the late observations carried the highest probability of detecting it. These distributions illustrate that even with precise knowledge of the host AGN properties, the predicted flare delay remains broadly distributed over timescales of months to years, underscoring the necessity of long-term, cadenced monitoring of AGNs within high-probability GW localization regions.

\section{Conclusion}\label{sec:Conclusion}

We conducted an extensive optical follow-up of the S240413p gravitational-wave event using the STEP program, including both early and late-time imaging and spectroscopic observations of two AGN-associated candidates, STEP2024gab and STEP2024phe. No clear electromagnetic counterpart was identified in the available photometric data.

Seasonal visibility constraints introduced a gap in the ZTF observed light curves, during which a potential flare could have occurred undetected. This highlights one of the key challenges in identifying electromagnetic counterparts to binary black hole mergers in AGN disks (dark flares). Current theoretical models \cite{juan_novo,Chen_2024,Tagawa2023b} predict that such dark flares may occur with delays ranging from $\sim$1 day up to $\sim$450 days after the merger, making continuous monitoring difficult. In addition, the merger may not have occurred within the AGN disk itself, or any emission could have been obscured by intrinsic AGN variability.

Another important limitation concerns the physical conditions required to launch relativistic jets from the kicked remnant black hole. Efficient jet formation likely requires a strongly magnetized environment. However, the magnetization properties of AGN disks remain poorly constrained. In particular, a low ionization fraction in the outer disk regions could decouple the gas from magnetic fields, suppressing magnetorotational instability and limiting magnetic field amplification, thereby hindering jet formation \citep{Chen_2024}. These considerations suggest that detectable dark flares require a combination of favorable physical conditions, implying that the fraction of BBH mergers in AGN disks producing observable electromagnetic counterparts may be significantly smaller than the total merger rate in such environments.

Current BBH--AGN models predict two broad classes of optical/UV counterparts. The first corresponds to early-type flares with short delays and durations ($\lesssim 50$ days), while the second consists of late-type flares with longer delays and extended durations ($\sim 50$--400 days). In addition to optical emission, multiwavelength counterparts may also arise across the electromagnetic spectrum, including gamma-ray \cite{zhang2025}, X-ray \cite{Kimura2021,Chen2024}, and radio bands \cite{Yi2019}. Radio emission in particular may resemble fast radio bursts but on longer characteristic timescales. Since emission in the infrared and hard X-ray bands is expected to remain relatively unobscured even within geometrically thick AGN disks \cite{Zhang2026}, coordinated multiwavelength observations represent a promising strategy for identifying these counterparts.

Future facilities will significantly improve the prospects for detecting such events. The Vera C. Rubin Observatory will enable deep, high-cadence monitoring of AGNs within high-probability gravitational-wave localization regions, increasing the likelihood of identifying long-duration flares. An optimal optical follow-up strategy for BBH mergers therefore combines rapid, deep observations in the days immediately following the merger to capture short-delay flares with longer-term monitoring through LSST. In practice, two complementary approaches appear most effective: systematic monitoring of AGNs located within the high-probability regions of gravitational-wave localizations and archival searches for flaring activity temporally coincident with merger events \cite{He_2025}.

Finally, characterizing potential host AGNs remains an important component of counterpart searches, although the large number of AGNs typically present within GW localization regions makes a complete characterization observationally impractical. Instead, prioritizing a subset of candidates based on readily accessible properties, such as SMBH mass estimates, accretion rate indicators, or archival variability, can help identify the most promising hosts. When combined with precise measurements of the remnant black hole mass provided by the LVK collaboration, these properties can constrain the expected flare delay and duration predicted by AGN-assisted merger models. Incorporating such physically motivated selection criteria into follow-up strategies will help focus observational resources on the most favorable candidates and improve the prospects for identifying electromagnetic counterparts to BBH mergers in AGN environments in future observing runs.

\section{Acknowledgements}

The authors made use of Sci-Mind servers machines developed by the CBPF AI LAB team and would like to thanks P. Russano, G. Teixeira and M. Portes de Albuquerque for all the support in infrastructure matters. We thank Murilo Marinello and Pedro Humire from S-PLUS collaboration for useful comments and discussions. C. D. Kilpatrick gratefully acknowledges support from the NSF through AST-2432037, the HST Guest Observer Program through HST-SNAP-17070 and HST-GO-17706, and from JWST Archival Research through JWST-AR-6241 and JWST-AR-5441. The ZTF forced-photometry service was funded under the Heising-Simons Foundation grant 12540303 (PI: Graham). This research uses services or data provided by the SPectra Analysis and Retrievable Catalog Lab (SPARCL) and the Astro Data Lab, which are both part of the Community Science and Data Center (CSDC) Program of NSF NOIRLab. NOIRLab is operated by the Association of Universities for Research in Astronomy (AURA), Inc. under a cooperative agreement with the U.S. National Science Foundation. Based in part on observations obtained at the Southern Astrophysical Research (SOAR) telescope, which is a joint project of the Minist\'{e}rio da Ci\^{e}ncia, Tecnologia e Inova\c{c}\~{o}es (MCTI/LNA) do Brasil, the US National Science Foundation's NOIRLab, the University of North Carolina at Chapel Hill (UNC), and Michigan State University (MSU). SP is supported by the international Gemini Observatory, a program of NSF NOIRLab, which is managed by the Association of Universities for Research in Astronomy (AURA) under a cooperative agreement with the U.S. National Science Foundation, on behalf of the Gemini partnership of Argentina, Brazil, Canada, Chile, the Republic of Korea, and the United States of America.

\bibliography{step}

\appendix{Appendix}

\end{document}